\documentclass[aps,prb,twocolumn,showpacs,superscriptaddress,longbibliography]{revtex4-1}  % for review and submission
\usepackage{graphicx}
\graphicspath{ {Figs/} }

\def\bibsection{\section*{\refname}}

\usepackage{times}

\usepackage{amsmath,amsthm}
\usepackage{amssymb}
\usepackage{booktabs}
\usepackage{color}
\usepackage{natbib}
\usepackage{bm, mathrsfs}
\usepackage{amsfonts}
\usepackage[scr=rsfso,cal=zapfc,frak=euler,bb=ams]{mathalfa}

\usepackage[utf8]{inputenc}
\usepackage[tight,sf]{subfigure}%

\newcommand{\be}{\begin{eqnarray}}
\newcommand{\ee}{\end{eqnarray}}

\newcommand{\fp}[1]{\textcolor{blue}{{#1}}}
\newcommand{\reply}[1]{\textcolor{black}{#1}}
\newcommand{\replyfinal}[1]{\textcolor{black}{#1}}

\usepackage{hyperref}
\hypersetup{colorlinks=true, citecolor=blue, urlcolor=blue, linkcolor=blue}

\usepackage{comment}

\begin{document}

\title{
Absence of diffusion in pilot-wave hydrodynamics: A classical analog of Anderson localization}

\title{
%Classical wave-particle analog of Anderson localization\\ 
%Classical wave-particle localization in disordered landscapes
%\\Wave-like localization of classical particles in random media 
Anderson localization of walking droplets}

\author{Abel J. Abraham}
\affiliation{Department of Mathematics, University of North Carolina, Chapel Hill, NC 27599, USA}

\author{Stepan Malkov}
\affiliation{Department of Mathematics, University of North Carolina, Chapel Hill, NC 27599, USA}

\author{Frane A. Ljubetic}
\affiliation{Department of Mathematics, University of North Carolina, Chapel Hill, NC 27599, USA}

\author{Matthew Durey}
\affiliation{School of Mathematics and Statistics, University of Glasgow, University Place, Glasgow, G12 8QQ, UK}

\author{Pedro J. S\'aenz}
\email[E-mail:~]{saenz@unc.edu}
\affiliation{Department of Mathematics, University of North Carolina, Chapel Hill, NC 27599, USA}

%% ---------- MATT'S COMMANDS ------------
\newcommand{\x}{\bm{x}}
\newcommand{\kb}{\bm{k}}
\newcommand{\yb}{\bm{y}}
\newcommand{\vb}{\bm{v}}
\newcommand{\pb}{\bm{p}}
\newcommand{\Ab}{\bm{A}}
\newcommand{\Bb}{\bm{B}}
\newcommand{\me}{\mathrm{e}}
\newcommand{\cb}{\bm{c}}
\newcommand{\DtN}{\mathscr{L}}
\newcommand{\FT}{\mathscr{F}}

\newcommand{\matt}[1]{\textcolor{red}{#1}}

\date{\today}
\begin{abstract}

Understanding the ability of particles to maneuver  through disordered environments is a central problem in innumerable settings, from active matter and biology to electronics.
Macroscopic particles 
ultimately exhibit diffusive motion when their energy exceeds the characteristic potential barrier of the random landscape. 
In stark contrast, 
wave-particle duality causes electrons in \reply{sufficiently} disordered media to come to rest even  when the potential is weak -- a remarkable phenomenon known as Anderson localization.
Here, we present a hydrodynamic active system with wave-particle features,  a millimetric droplet self-guided  by its own wave field over a submerged random topography, whose dynamics exhibits localized statistics analogous to those of electronic systems. Consideration of an ensemble of particle trajectories reveals a suppression of diffusion when the guiding wave field extends over the disordered topography. We rationalize mechanistically the emergent statistics by virtue of the  wave-mediated resonant coupling between the droplet and topography, which produces an attractive wave potential about the localization region. 
This hydrodynamic analog, which demonstrates how a classical particle may localize like a wave,  suggests new directions for future research in various areas, including active matter, wave localization, many-body localization, and  topological matter.

\begin{comment}
    
% SCIENCE (125 words, no references)
Macroscopic particles in disordered environments ultimately exhibit diffusive motion when their energy exceeds the characteristic potential barrier of the heterogeneous background. In contrast, subatomic particles in random media come to rest even when the disorder is weak due to their wave-particle duality, an intriguing phenomenon known as Anderson localization so far thought exclusive to the quantum realm. 
We present a macroscopic wave-particle system,  a millimetric droplet guided by its own wave field over a submerged random topography, that exhibits an absence of diffusion analogous to that of quantum systems. The emergent localized statistics are rationalized by virtue of the resonant coupling between the droplet and its self-excited guiding wave field. This hydrodynamic wave-particle analog suggests new directions for future research in various areas, including topological matter and wave localization.

\end{comment}

\end{abstract}

\maketitle

\section{Introduction}

\begin{figure*}
\includegraphics[scale=1]{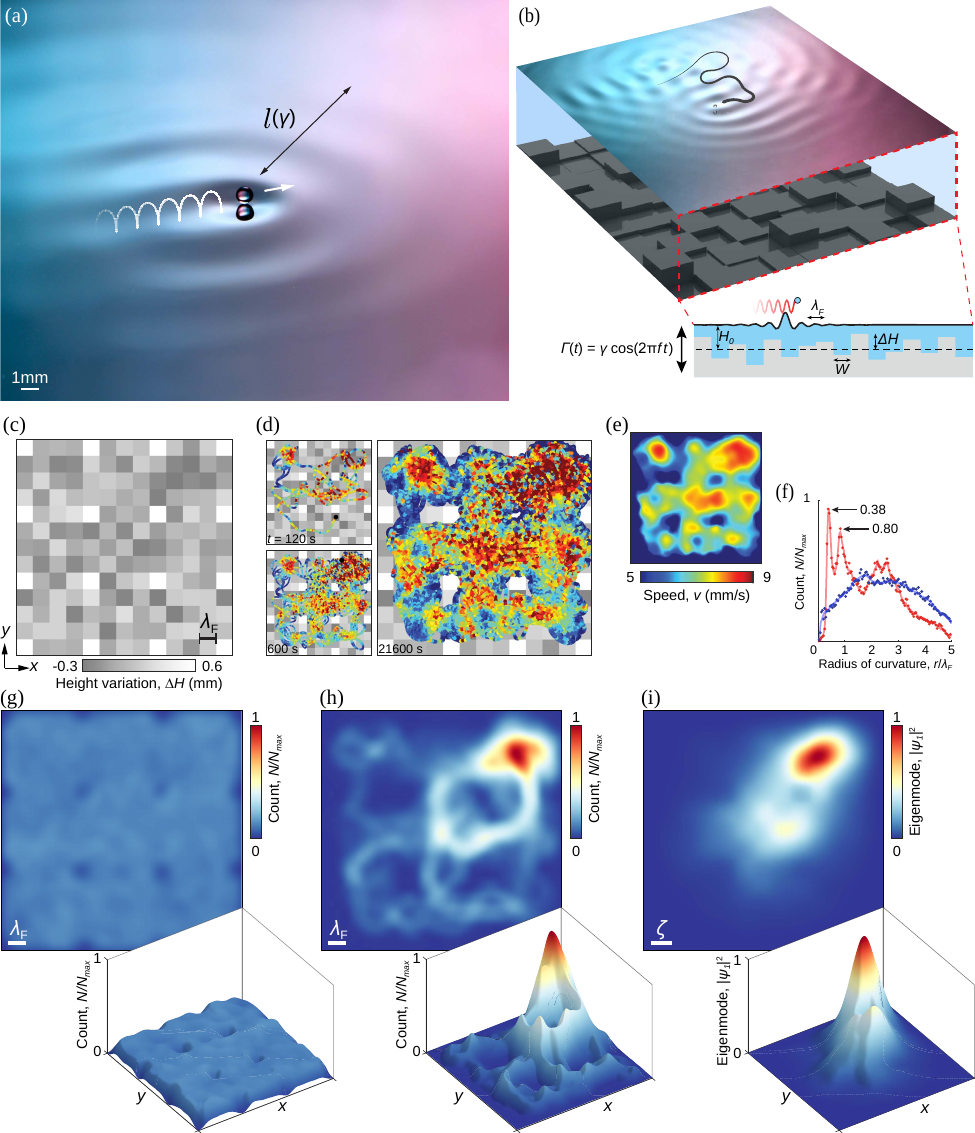}
\caption{\textbf{Localization of walking droplets over submerged random topographies.} 
(a) Oblique view of a walking droplet self-propelled by its own guiding wave field.
%, whose spatial extent, $l(\gamma)$, increases with the bath's forcing acceleration, $\gamma$.
(b) In our experiments, the droplet moves erratically (Supplementary Video 1) due to the nonlocal influence of a submerged disordered topography (not to scale vertically) composed of square tiles, each of width $W=5.25$\,mm  (matching the Faraday wavelength, $\lambda_\mathrm{F}$) and random height  (uniform distribution in the interval $\pm\Delta H$) around a base depth $H_0=1.85$\,mm.
(c) Random realization of the submerged topography used in our experiments. 
(d) At high memory ($\gamma/\gamma_\mathrm{F}=99.8\,\%$), the walker trajectory (colored according to the instantaneous speed) reveals a chaotic exploration of the domain.
(e) After 6\,h, the average local speed shows  regions of higher and lower speeds roughly corresponding to deeper and shallower areas.
(f) Comparison of the radius of curvature of the trajectory for our walker (red) and a `waveless' particle (blue) in the same random potential (each normalized by the overall maximum, $N_\mathrm{max}$). The walker moves at an average speed $v_0=7.3$\,mm/s, turning along preferred radii of curvature, $r/\lambda_\mathrm{F}\sim 0.38$ and 0.80. 
The position histogram for the waveless particle is uniform (g). Notably, the walker position histogram exhibits a localization region (h), which bears a strong resemblance to the first eigenmode %(normalized by the maximum amplitude) 
of Schr\"odinger's equation for the same potential in the weak regime (i).
} 
\label{fig:main_Fig_1}
\end{figure*}

Disorder is omnipresent in nature. Investigating the motion of particles in heterogeneous media has thus always captivated scientists across disciplines, yielding an ever-growing list of transformative breakthroughs.
%that have reshaped our understanding of the world around us. 
Modern examples are abundant in the fields of active matter\cite{2013Marchetti_Review,bechinger2016active}, including systems of flocking colloids cruising over random obstacles\cite{morin2017distortion}  or microbial dispersion in porous media \cite{deAnna2021}, and condensed matter, including states of matter insensitive to defects\cite{hasan2010topological} or particle-like magnetic excitations in 
 quenched disorder\cite{reichhardt2022statics} that show promise for the development of spintronics\cite{wiesendanger2016nanoscale,Dieny2020}.
Classic examples comprise the 
erratic motion of pollen grains caused by the molecules of their enveloping fluid\cite{Einstein1905}, and the turbulent transport of plankton and other marine ecosystems\cite{levy2018the}. 
%Recent examples comprise microbial dispersion in porous media \cite{deAnna2021}, or the discovery of states of matter insensitive to defects\cite{hasan2010topological}, and particle-like magnetic excitation in quenched disorder\cite{reichhardt2022statics} that show promise for the development of spintronics\cite{wiesendanger2016nanoscale,Dieny2020}.
%
Of particular interest, for both fundamental and practical reasons, is the comparative study of classical and subatomic particle dynamics in random landscapes. When macroscopic particles, such as billiard balls, move with high kinetic energy across a weak disordered background, they are deflected by small random forces that eventually cause their motion to become diffusive in two (or higher) dimensions \cite{bechinger2016active}.
Subatomic particles display a fundamentally different behavior. An electron  
%in a metal with impurities 
may spontaneously come to a halt, or `localize', over a relatively weak random potential  for a sufficient degree of disorder\cite{anderson1958Absence,anderson2008anderson}; this phenomenon, Anderson localization, has reshaped condensed matter physics\cite{Lagendijk2009Fifty,Kramer_1993,imada1998metal}.
Here, we demonstrate that macroscopic  `walking droplets'\cite{Couder2005a} (Fig.\,\ref{fig:main_Fig_1}\fp{a}), which propel along the surface of a vibrating fluid bath guided by their own self-excited wave field, may exhibit analogous localization.

%that hinders the conductance of electrons. %in metals with impurities. 

Anderson localization results from the wave-like behavior of electrons\cite{Lagendijk2009Fifty,Kramer_1993,imada1998metal}. Hence, significant research efforts have pursued  demonstrating the universality\cite{Kramer_1993} of wave localization across diverse systems, including matter waves\cite{billy2008direct,roati2008Anderson,Jendrzejewski2012}, microwaves and light waves \cite{dalichaouch1991microwave,wiersma1997localization,Schwartz2007Transport,segev2013anderson,Yamilov2023}, ultrasound\cite{hu2008localization}
 and  water waves\cite{belzons1988gravity,devillard1988Localization}.  
 %All of these being waves systems, 
 Today, Anderson localization is  thus widely regarded as a purely classical wave phenomenon\cite{filoche2012universal}. Inspired by their quantum counterparts, however, our interest is to investigate whether classical \textit{particles} may localize like waves.

To test this hypothesis, we considered the wave-mediated spontaneous motion of a bouncing droplet on the surface of a vibrating fluid bath (Fig.\,\ref{fig:main_Fig_1}\fp{a}) subjected to vertical forcing acceleration $\Gamma(t)=\gamma\cos(2\pi f t)$, where $\gamma$ is the maximum acceleration, $f$ the oscillation frequency, and $t$ time. Despite being macroscopic  objects, these walking droplets display dual wave-particle behaviors reminiscent of those arising in quantum systems\cite{Bush2021Hydrodynamic}. 
%Coalescence is prevented by the intervening air layer between the droplet and the bath \cite{Couder2005b}. 
As the forcing acceleration is increased, the 
%droplet's vertical motion undergoes a spontaneous period-doubling cascade\cite{Protiere2006,Molacek2013a}. The 
droplet begins to bounce at half of the driving frequency, thus achieving resonance with the most unstable wave mode of the bath\cite{Protiere2006,Molacek2013b}. At each bounce, the droplet excites decaying circular waves, with wavelength $\lambda_\mathrm{F}$, whose 
influence on the droplet dynamics
%longevity and spatial extent, $l(\gamma)$ (Fig.\,\ref{fig:main_Fig_1}\fp{a}), 
increases with the forcing acceleration\cite{Eddi2011}.
When the driving exceeds  a critical walking threshold, $\gamma>\gamma_\mathrm{w}$, vertical bouncing becomes unstable and the droplet starts to land on the slope of the wave field generated at previous impacts. The droplet thus experiences a wave-induced horizontal force, $\mathbf{F}(\bm{x},t)\propto -\nabla\eta|_{\bm{x}=\bm{x}_\mathrm{p}}$, proportional the slope of the underlying wave field, $\eta(\bm{x},t)$, at the particle's location, $\bm{x}_\mathrm{p}(t)$\cite{Protiere2006,Eddi2011}. %Walking droplets, or `walkers', in a homogeneous bath thus move along straight paths at constant speeds, acting as self-propelled wave sources\cite{Bush2021Hydrodynamic}.
%Walking droplets, or `walkers', thus act as self-propelled wave sources\cite{Bush2021Hydrodynamic}. In a homogeneous bath, walkers move along straight paths at constant speeds\cite{Bush2021Hydrodynamic}.
Acting as self-propelled wave sources\cite{Bush2021Hydrodynamic}, walking droplets, or `walkers',  thus move along straight paths at constant speeds \reply{on the surface of}  a homogeneous bath\cite{Protiere2006}.

To investigate the walker motion in disordered media, we leveraged variable bottom topography 
%as a versatile strategy 
to subject the droplet to spatially-varying  %external/background? 
potentials\cite{Saenz2018,Saenz2019,Saenz2021}.
In our experiments, the bath's bottom profile was composed of square tiles, each with a random height around a base depth $H_0=1.85$\,mm drawn from the uniform distribution in the interval $\Delta H=\pm0.3$\,mm 
(Fig.\,\ref{fig:main_Fig_1}\fp{b}). Hence, the fluid depth is $H(\bm{x})=H_0 - \Delta H(\bm{x})$. 
 A wave damper surrounding the heterogeneous bottom  prevented the droplet from escaping the disordered domain (\ref{sec:SI_experiment}, Experiments).
The tile width, $W=5.25$\,mm, was selected to match the characteristic Faraday wavelength, $\lambda_\mathrm{F}$, dictated by the capillary-gravity dispersion relation\cite{Kumar1994Parametric} for the base depth, $H_0$. Notably, the depth variations in our experiments have a negligible influence on $\lambda_\mathrm{F}$. 
The droplet, with radius $R=0.368$\,mm, and  fluid bath were composed of the same  20 cSt silicon oil with kinematic viscosity $\nu$, density $\rho$ and surface tension $\sigma$. The bath was vibrated with an electromagnetic shaker with oscillation frequency $f=70$\,Hz (Fig.\,\ref{fig:sup_fig_schematic}).
The experimental topography (Fig.\,\ref{fig:main_Fig_1}\fp{c}) included regularly spaced pillars, $\Delta H=0.6$\,mm, to facilitate  data acquisition, which play no significant role in the localization (\ref{sec:SI_simulations}, Simulations).
Equipped with this 
%versatile  
strategy to tune spatial disorder in the walker system, we investigated the droplet's  motion over submerged random topographies.

The quantum-like effects of the walker system emerge when the droplet motion is strongly influenced by the underlying wave field\cite{Bush2021Hydrodynamic}, which is characterized by a spatial extent, $l$, and a  decay, or `memory', time, $T_\mathrm{M}$ \cite{Eddi2011,tadrist2018Faraday}. 
Both of these spatio-temporal features increase as the forcing acceleration approaches the Faraday threshold, $\gamma_\mathrm{F}=3.775g$ ($g=9.81\,\mathrm{m/s}^2$),  at which 
the entire bath surface becomes unstable to standing waves\cite{Kumar1994Parametric}. We thus focused on the walker's statistical behavior in the high-memory regime, $\gamma/\gamma_\mathrm{F}=99.8\,\%$, when the waves are  long-lived, $T_\mathrm{M}\sim 26$\,s, and span a  large area of the bath, $l\sim \reply{7}\lambda_\mathrm{F}$. No waves exist in the absence of the droplet ($\gamma<\gamma_\mathrm{F}$).
To achieve statistical significance, we recorded the droplet motion for 6\,h, which corresponded to  756,000 bounces and roughly 2,300 domain crossings. The data was acquired in 20\,min intervals to maintain the prescribed memory (\ref{sec:SI_experiment}, Experiments).
%, which may drift slowly owing to variations in surface tension and viscosity caused by changes in the ambient temperature (\ref{sec:SI_experiment}, Experiments).

\vspace{4pt}
\section{Localized statistics}

The walking droplet moved erratically above the heterogeneous bottom topography  due to  deformations in the wave field produced by the disordered features below the bath surface (Fig.\,\ref{fig:main_Fig_1}\fp{d}, Supplementary Video 1). The chaotic trajectory is characterized by random  speed fluctuations that typically occur at a length scale shorter than that of the underlying topography, $W$, thus evidencing the presence of wave-mediated forces produced by distant submerged features. The spatially-averaged speed map revealed  areas of higher and lower speeds roughly corresponding to deeper and shallower regions, respectively (Fig.\,\ref{fig:main_Fig_1}\fp{e}),  indicating a depth-dependent speed envelope superimposed on the high-frequency speed fluctuations. %Moreover, the speed distribution (Fig.\,\ref{fig:main_Fig_1}\fp{f}) concentrated around an average speed, $v_0=7.3$\,mm/s, which roughly corresponds to droplet's free-walking speed\cite{Eddi2011,Molacek2013b} in a homogeneous bath with the same average depth.

To contextualize the distinct walker behavior, we compared their statistics to those arising from a `waveless' particle that evolves classically according to $m \ddot{\bm{x}}_\mathrm{p}(t)=-\nabla V(\bm{x}_\mathrm{p})$, and a quantum particle governed by Schr\"odinger's equation, $\mathrm{i}  \hbar \Psi_t (\bm{x},t)=-\hbar^2 \nabla^2 \Psi/2m+V(\bm{x}) \Psi$, where $m$ is the particle mass and $\Psi$ the wave function. The particles were subject to a random potential of the same form as the experimental bottom topography, $V(\bm{x})=-\kappa ( H(\bm{x})-\max H(\bm{x}) )$, up to a re-scaling constant, $\kappa$. 
Anderson localization is unique because the particles become effectively trapped \reply{when the potential is sufficiently disordered but weak relative to the particle's kinetic energy}\cite{MullerDelande2011Andersontheory} 
% $E_\zeta E/V_0^2\gg 1$ and $E_\zeta/V_0\gg 1$. Here, $E$ is the total energy of the particle, $E_\zeta$ the energy associated to the potential's correlation length $\zeta$, and $V_0$ the characteristic potential energy 
(\ref{sec:SI_quantum_localization}, Quantum Localization).
In our simulations, we thus adjusted the  constant $\kappa$ and the initial particle energy to ensure that the total energy was larger than the background potential. For the waveless particles, we also matched the average speed to that in our experiments.  
%For the quantum particles, we maintained $E_\zeta E/V_0^2> 120$ and $E_\zeta/V_0> 100$ throughout this study.
The effect of the wave damper along the border of the heterogeneous region was modeled through a confining potential  for the waveless particle, and Dirichlet boundary conditions  in the quantum simulations (\ref{sec:SI_simulations}, Simulations, 
\ref{sec:SI_quantum_localization}, Quantum Localization).

The emergent speed statistics for the walker and waveless particle are similar (Fig.\,\ref{fig:sup_fig_waveless_experiment_simulations}, Fig.\,\ref{fig:sup_curvature_maps}\fp{a}). However, the distribution of the radius of curvature of the walker trajectory revealed a lack of tight turns and, instead, significant peaks at $r/\lambda_\mathrm{F}\sim 0.38$ and 0.80 (Fig.\,\ref{fig:main_Fig_1}\fp{f}), roughly corresponding to the preferred radii observed in experiments involving walkers forced along curved paths\cite{Fort2010}. The trajectory of the waveless particle exhibited more violent velocity reversals and wider turns (Fig.\,\ref{fig:main_Fig_1}\fp{f}, Fig.\,\ref{fig:sup_fig_waveless_experiment_simulations}\fp{a}).
The walker thus turns along more restricted trajectories due to the influence of the pilot wave field, which sets a kinematic constraint on the preferred path curvature\cite{Blitstein2024}.

A much more profound difference is observed when comparing the spatial distributions of the particles. The position histogram for the waveless particle is relatively homogeneous (Fig.\,\ref{fig:main_Fig_1}\fp{g}); the particle thus explores the domain uniformly.
 %, showing no preference for any particular location. 
In contrast, the walker's position histogram  features a prominent peak, or `localization' region (Fig.\,\ref{fig:main_Fig_1}\fp{h}),  where the walker returns more frequently and spends more time.
Notably, the observed  walker localization bears a strong resemblance to the first %exponentially-decaying 
eigenmode in the Anderson regime for a quantum particle in the same random potential (Fig.\,\ref{fig:main_Fig_1}\fp{i}, see \ref{sec:SI_quantum_localization}, Quantum localization).

\begin{figure*}
\includegraphics[scale=0.9]{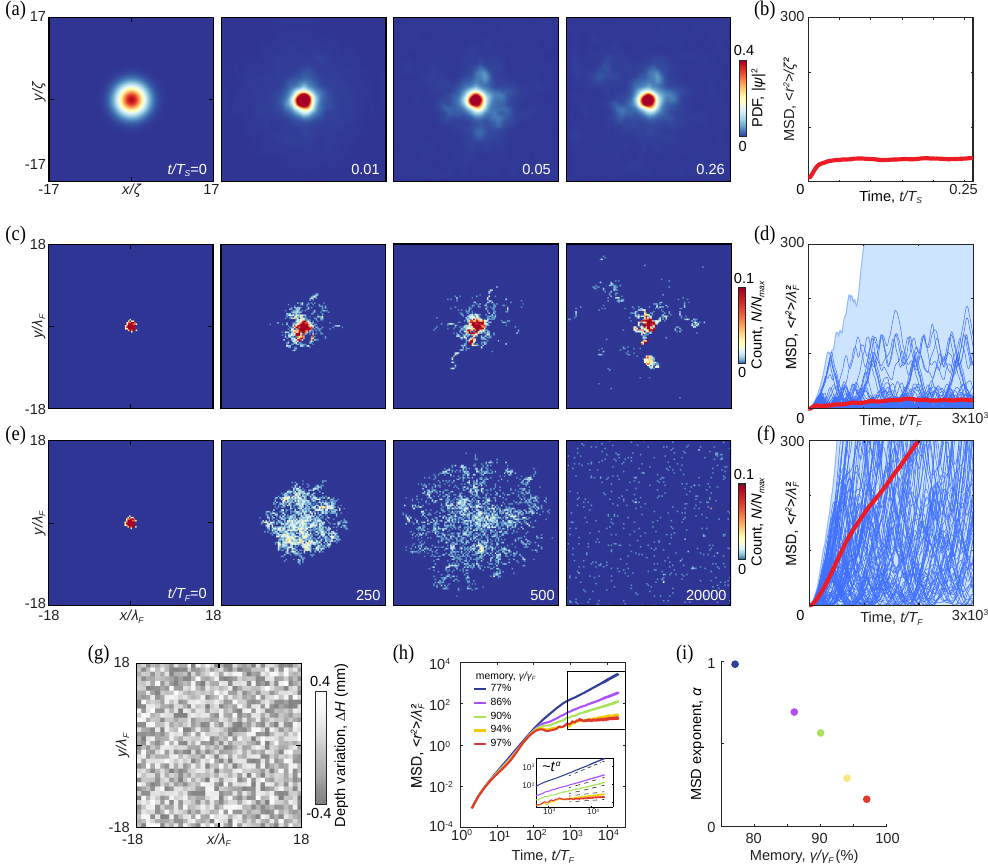}
\caption{\textbf{Absence of diffusion in the walker system.} (a) Anderson localization is observed in the  temporal evolution of a quantum wave packet, $|\Psi|^2$, over a weak random potential.
The probability-density function becomes localized due to the disorder; the distribution's mean-squared displacement (MSD)  saturates in the long-time limit (b).
%, even when the potential is weak relative to the particle's energy. 
%(b) The average mean-squared displacement saturates in the long-time limit indicating that the particle's dispersion ceases spontaneously.
%in a regime where a classical particle subject to local forces would evolve diffusely.
(c-d) A similar absence of diffusion is observed in simulations of an ensemble of walking droplets at high memory, $\gamma/\gamma_\mathrm{F}=97\,\%$, which become localized  due to the influence of their wave field. (e-f) At low memory, $\gamma/\gamma_\mathrm{F}=77\,\%$, when the walker's wave field is smaller and shorter-lived, the droplets move freely.
 (g) Sample of random topography.
 (h) Long-time evolution of the MSD shown in (d, f) and from other values of memory. 
 (i) Dependence on memory of the MSD's temporal exponent, showing a gradual decrease as the memory is increased. 
 %(i) Speed, and (j) curvature distributions corresponding the simulations shown in (g). 
 The average \reply{walker} speed is \reply{$v_0=5.1$}\,mm/s \reply{in all cases}.
  In (a-b), we averaged over 20 random  potentials. Space and time are normalized according to the correlation length, $\zeta$, and $T_\mathrm{S}=mL^2/\hbar$, respectively
(\ref{sec:SI_quantum_localization}, Quantum localization). 
In (c-j), we averaged over \reply{8} random topographies with $H_0=1.45$\,mm and \reply{400} walker trajectories for each realization. Time is normalized by the Faraday period, $T_\mathrm{F}=2/f$ (\ref{sec:SI_simulations}, Simulations). See Supplementary Videos 2 for a video animations of (a, c, e).
%is observed in the simulation of an ensemble of walkers
%A similar behavior is observed in the simulation of an ensemble of walkers 
%owing to the influence of the wave field, which extends a relatively large region of the random medium
%
%Simulation of Ensembles of 1000 walkers for 
%\markblue{For quantum, (A), average values are $\langle E\rangle = 16.61\pm 1.69$, $\langle E_\zeta\rangle = 1294.64\pm 36.46$, $\langle V_0\rangle = 13.19\pm 0.15$, $\langle \zeta\rangle = 0.0278\pm 3.96\times 10^{-4}$, and $E_\zeta E/V_0^2=123.6$}
\label{fig:main_Fig_2}}
\end{figure*}

To further probe the mechanisms underpinning the walker localization, we complemented our experiments with simulations of walking droplets with a quasi-potential fluid model\cite{Faria2016}, for which the bath surface elevation evolves according to
\begin{equation}
\begin{aligned}
u_t &= -G(t)\eta + \frac{\sigma}{\rho} \Delta \eta + 2\nu \Delta u - \frac{P}{\rho}, \\
\eta_t &= \DtN u + 2\nu \Delta \eta, 
\label{eq:wavesmodel} 
\end{aligned}
\end{equation}
where $u(\x,t)$ is the linearized free-surface velocity potential, $G(t)$ the effective gravitational acceleration, and $P$ the pressure exerted on the free surface during each bounce. The operator  $\DtN(\x)=-\nabla \cdot b(\x)\nabla$ captures the influence of variable bottom topography by adjusting the local wave speed through an effective depth, $b(\x)$. The droplet position evolves according to $m \ddot{\bm{x}}_\mathrm{p}(t)+D(t)\dot{\bm{x}}_\mathrm{p}(t)=-F(t)\nabla \eta(\bm{x}_\mathrm{p},t)$, where $D(t)$ is the viscous dissipation and $F(t)$ the interaction force (\ref{sec:SI_theory}, Theory).  
Armed with this walker model, which provides good agreement with our experiments (Fig.\,\ref{fig:sup_fig_waveless_experiment_simulations}\fp{b-c}), we examined next whether the walker localized statistics suppress diffusion as in the quantum system.

\vspace{4pt}
\section{Supression of diffusion}

In Anderson localization, the evolution of a quantum particle,  represented by the probability density function (PDF), $|\Psi|^2$, spontaneously comes to a halt over a sufficiently random potential even though the particle's energy is larger than the background potential (Fig.\,\ref{fig:main_Fig_2}\fp{a}). After an initial ballistic phase caused by short-time scattering\cite{MullerDelande2011Andersontheory}, the PDF's mean-squared displacement (MSD), $\langle r^2(t)\rangle=\iint_\Omega|\Psi|^2|\bm{x}|^2 \,\mathrm{d}S$, saturates in time  (Fig.\,\ref{fig:main_Fig_2}\fp{b}, \ref{sec:SI_quantum_localization}, Quantum localization). To test whether a similar suppression of diffusion occurs in our hydrodynamic system, we performed walker simulations with \reply{8 random}  realizations of the topography \reply{over domains larger than those feasible in  experiments (Fig.\,\ref{fig:main_Fig_2}\fp{g}). Each realization was} explored by \reply{400} walkers for \reply{$2\times10^4$} Faraday periods, $T_\mathrm{F}=2/f$.
%initialized with normally distributed position and momentum. 
We tuned the base depth, $H_0=1.45$\,mm, and particle's average speed, \reply{$v_0=5.2$}\,mm/s. (\ref{sec:SI_simulations}, Simulations).
Indeed, at high memory, $\gamma/\gamma_\mathrm{F}=97\,\%$, when wave-like behavior of walking droplets is most prominent\cite{Bush2021Hydrodynamic}, we observed that the position histogram remained localized and effectively froze (Fig.\,\ref{fig:main_Fig_2}\fp{c}). Similarly, the MSD of the particle ensemble, $\langle r^2(t)\rangle=\frac{1}{N}\sum _{i=1}^{N}|\x_{\mathrm{p}_i}|^{2}$, saturated in  the long-time limit, revealing an absence of diffusion in the walker system (Fig.\,\ref{fig:main_Fig_2}\fp{d}).

\begin{figure*}
\includegraphics[scale=0.9]{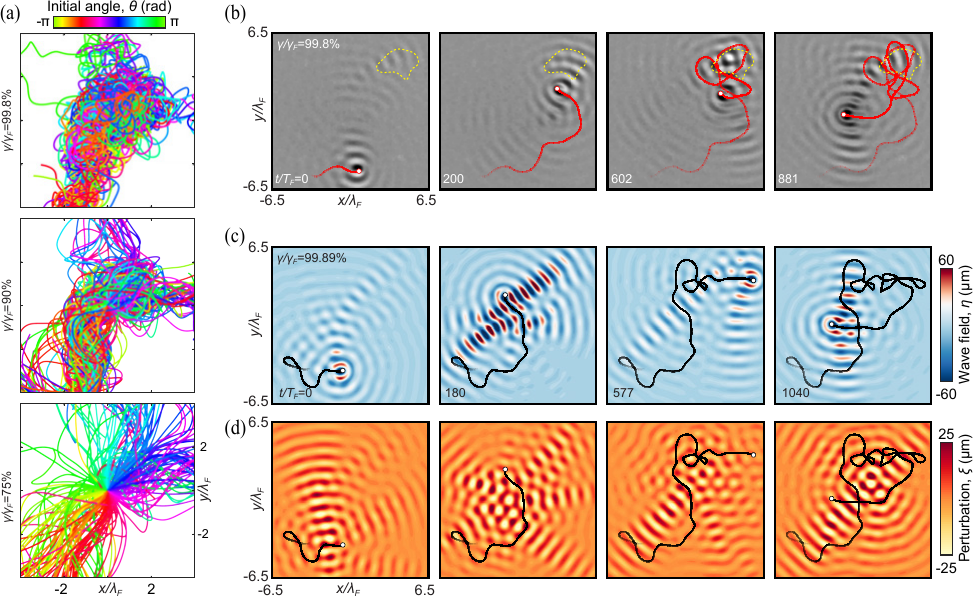}
\caption{\textbf{Spatio-temporal non-local interactions  cause the walker localization.} 
(a)  The initial spreading of the simulations described in Fig.\,\ref{fig:main_Fig_2}\fp{c-i} reveal that the droplet trajectories are characterized by frequent turns and loops at higher memory, when walker localization is observed, but become more rectilinear at low memory, when the walker motion is more diffusive {(Supplementary Video 2)}.
\reply{For illustrative purposes, here we increased the memory up to 99.8\%, which is the experimental value but higher than what is numerically feasible for the long-time statistics shown in  Fig.\,\ref{fig:main_Fig_2}\fp{h-i}.} 
(b) Visualization of the wave field in our experiments reveals the emergence of beams connecting the droplet and localization region (yellow dashed line) that draw the droplet in. Once inside the localization region, the superposition of slowly-decaying waves traps the droplet in loops that significantly increase the trapping time.
(c) Similar wave-mediated interactions are observed in our walker simulations.
(d) The attracting beams, and trapping waves are more apparent in the topography-induced wave perturbation, $\xi = \eta - \bar{\eta}$, obtained by subtracting from the wave field, $\eta$, shown in (c) the computed wave field, $\bar{\eta}$, of a droplet following the same trajectory in homogeneous bath of the same base depth, $H_0$.  
\reply{The wave fields in (b,c) are strobbed at the Faraday frequency with a phase at which the waves are near their maximum amplitude.
}
See Supplementary Video 3 for video animations of (b-d)\reply{.}}
\label{fig:main_Fig_3}
\end{figure*}

To demonstrate that the droplet localization is correlated to the spatio-temporal attributes of its pilot wave, we also examined the walker dynamics at low memory, $\gamma/\gamma_\mathrm{F}=77\,\%$, when the wave field is less extensive, $l\sim2\lambda_\mathrm{F}$, and decays quickly, $T_\mathrm{M}\sim 0.23$\,s. The particle's average speed was maintained by readjusting the impact phase (\ref{sec:SI_simulations}, Simulations). In this regime, droplets move freely throughout the heterogeneous media (Fig.\,\ref{fig:main_Fig_2}\fp{e}-\fp{f}), indicating a diminished influence of the submerged topography on the wave field. As the memory decreases, the walker dynamics thus approaches the diffusive behavior of the waveless particle.
%that respond to local forces.
%more diffusive behavior, similar to that observed with the waveless particle, thus emerge in the walker system as the memory decreases.
\reply{We performed analogous simulations over a range of memories and quantified the walker's diffusive behavior through}
%Diffusive processes may be quantified by 
the MSD's asymptotic dependence on time\cite{bechinger2016active}, $\langle r^2(t)\rangle\sim t^\alpha$ (Fig.\,\ref{fig:main_Fig_2}\fp{h}). As the memory was increased, we observed a progressive suppression of diffusion, from a conducting  state ($\alpha\to 1$) at low memory to a localized  state ($\alpha\to 0$) when the forcing acceleration is near the Faraday threshold (Fig.\,\ref{fig:main_Fig_2}\fp{i}). 
%Ordinary diffusion ($\alpha= 1$) is not completely recovered for low forcing due to the finite size of the wave field, which remains significant relative to the topography's characteristic length, $W$, at the lowest memory that permits walking. 
%, $\gamma/\gamma_\mathrm{F}=83\,\%$. 
%Analysis of the speed and curvature distributions (Fig.\,\ref{fig:main_Fig_2}\fp{i}-\fp{j}) further confirmed the distinct behavior of walkers at high memory, which exhibit frequent turns with a preferred radius of curvature, and walkers at low memory, whose dynamics is closer to that of a waveless particle  (Fig.\,\ref{fig:main_Fig_1}\fp{f}-\fp{g}). 
%The absence of larger preferred radii of curvature is caused by the heightened influence of the topography in the shallower layer (Supplementary Figure 3).

\begin{figure}
\includegraphics[scale=1]{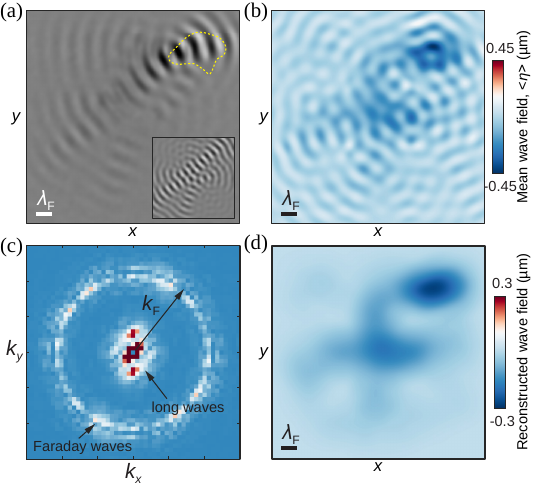}
\caption{\textbf{Superposition of long-wave modes rationalizes the mean wave field.} 
(a) Experimental Faraday waves observed at threshold, $\gamma=\gamma_\mathrm{F}$, emanating from the localization area  (yellow dashed line), which agrees with the theoretical Faraday mode (inset). 
(b) The mean (time-averaged) pilot wave, $\langle\eta\rangle$, computed from simulations matching the experiment (Fig.\,\ref{fig:sup_fig_waveless_experiment_simulations}\fp{c}), shows a large depression at the localization region. 
(c) The spectrum of the mean wave is composed of subcritical Faraday modes  and long waves, which may be obtained theoretically from the walker wave model \eqref{eq:wavesmodel}.
(d) The same large-scale envelope, which acts as an effective potential, \reply{is reconstructed from the position histogram and long-wave modes according to \eqref{eq:convolution}}.
}
\label{fig:main_Fig_4}
\end{figure}

\section{Wave-mediated mechanism}

%The walker trajectories exhibit frequent turns and loops at high memory, but become more rectilinear when the forcing is reduced  (Fig.\,\ref{fig:main_Fig_3}\fp{a}).
To rationalize the walker dynamics that result in localized statistics\reply{, which exhibits frequent turns with a preferred radius of curvature as the memory increases}
(Fig.\,\ref{fig:main_Fig_3}\fp{a}),  we investigated the  droplet-topography interaction mechanism through examination of the wave field. At any instant, the wave field is complex, the result of a superposition of waves created by the droplet's previous bounces\cite{Saenz2018}. Yet, both experiments (Fig.\,\ref{fig:main_Fig_3}\fp{b}) and simulations (Fig.\,\ref{fig:main_Fig_3}\fp{c})  showed two distinct wave-mediated resonant effects responsible for localization.  
%giving rise to the localized statics
 These effects become more evident in the topography-induced, or `anomalous'\cite{Saenz2019}, wave field, $\xi(\bm{x},t)=\eta-\bar{\eta}$ (Fig.\,\ref{fig:main_Fig_3}\fp{d}), obtained by subtracting from the simulated wave field, $\eta$ (Fig.\,\ref{fig:main_Fig_3}\fp{c}), that of the droplet following the same path in a homogeneous layer of the same base depth, $\bar{\eta}$. First, when the localization region is within range of the walker  wave field, beam-like waves\cite{Saenz2019} connecting the droplet and the localization region draw the droplet in. Second, once the droplet reaches the localization region, relatively larger waves are excited, which force the droplet to execute multiple loops, thus increasing the trapping time (Fig.\,\ref{fig:main_Fig_3}\fp{d}). These wave-mediated effects thus produce spatio-temporally non-local forces\cite{Blitstein2024} on the droplet that result in a peak in the position histogram. %\todo{Notably, the walker speed is larger in the localization region (Fig.\,\ref{fig:main_Fig_1}\fp{e}), opposite to the classical behavior of particles, which move more slowly in regions where they spend more time.} 
 %Similar spatio-temporal non-local effects were observed in the wave-mediated interaction of a walker and a submerged well\cite{Saenz2019}.
At low memory, when the walker motion is more diffusive, the relative extent and magnitude of the attracting beams and trapping waves are significantly  diminished 
(Fig.\,\ref{fig:sup_fig_LMwave}).

To understand the resonant waves that emanate from the localization region, which acts as an area of high excitability, we visualized experimentally the hydrostatic free surface at the instability threshold, $\gamma=\gamma_\mathrm{F}$, in the absence of the droplet (Fig.\,\ref{fig:main_Fig_4}\fp{a}, Supplementary Video 3). 
The localization region corresponds to the location where the first Faraday waves are observed, which bear a resemblance to the beam-like waves that are occasionally  excited in the anomalous walker wave field (Fig.\,\ref{fig:main_Fig_3}\fp{d}). 
%[\todo{Remove the rest of this paragraph.}] Notably, we also observed a correspondence between the form of the Faraday pattern and that of the mean pilot wave\cite{Saenz2018}, $\langle\eta\rangle$, which is characterized by a relatively wide depression centered at the localization region (Fig.\,\ref{fig:main_Fig_4}\fp{b}). To rationalize the form of the mean wave field, we investigated theoretically its spectrum (Fig.\,\ref{fig:main_Fig_4}\fp{c}), which is comprised of near-critical standing Faraday modes (Fig.\,\ref{fig:main_Fig_4}\fp{a}, inset), and long travelling waves emanating from each droplet impact. Through a perturbative Floquet analysis, we computed the slowest decaying wave modes of the walker model \eqref{eq:wavesmodel},  and, under the assumption of ergodicity \cite{DMB2018, DMW2020}, used the droplet's histogram to determine the mode weighting (\ref{sec:SI_theory}, Theory). 
%While unimportant for pilot-wave dynamics in a homogeneous bath,
%%Unlike the pilot wave in a homogeneous bath, 
%the resulting convolved wave field (Fig.\,\ref{fig:main_Fig_4}\fp{d})  indicates that long waves\cite{MGNB2015} play a significant role in the walker localization in disordered media, producing an effective potential about the localization region.

% ----------------------
% MEAN WAVE FIELD PARAGRAPH

\reply{We also observed a correspondence between the droplet histogram and the mean pilot wave\cite{Saenz2018}, $\langle \eta \rangle$, which is characterized by the superposition of Faraday waves about a relatively wide depression centered at the localization region (Fig.\,\ref{fig:main_Fig_4}\fp{b}) \cite{DMW2020}. The spectrum of the mean wave field (Fig.\,\ref{fig:main_Fig_4}\fp{c}) highlights a separation of length scales, consisting of near-critical standing Faraday waves and long travelling waves emanating from each droplet impact\cite{ESMFRC2011}. Under the assumptions of ergodicity and wave linearity, we may rationalize the form of the mean wave field in terms of the droplet histogram, $p_s(\x)$, via the relationship\cite{DMB2018, DMW2020}
\begin{equation}
\label{eq:convolution}
\langle \eta \rangle (\x) = \iint_{\mathcal{D}} p_s(\x') \eta_B(\x,\x')\,\mathrm{d}\x',
\end{equation}
where $\eta_B(\x,\x_{\mathrm{p}})$ is the wave field  generated by a bouncer at position $\x_{\mathrm{p}}$ in the domain $\mathcal{D}$. By decomposing $\eta_B$ into its long and Faraday wave components, which may be approximated through the analysis of the quasi-potential waves model \eqref{eq:wavesmodel}  (\ref{sec:SI_theory}, Theory), we used \eqref{eq:convolution} to numerically reconstruct the mean wave field separately, and so discern the features produced by each type of mode. While unimportant for pilot-wave dynamics in a homogeneous bath, the reconstructed mean wave field using the long-wave modes indicates that the complex reflection and transmission of long travelling waves over disordered media play a significant role in the walker localization, producing an effective long-range potential about the localization region that contributes to the sustained trapping of the droplet (Fig.\,\ref{fig:main_Fig_4}\fp{d}) \cite{DMW2020}. Notably, this bowl-shaped potential cannot be generated solely by the Faraday modes, whose elevation instead approximately vanishes when averaged over each wavelength.
Furthermore, a perturbative Floquet analysis of the system's Faraday modes demonstrates that the mean Faraday waves are of a highly complex form, and do not closely resemble any of the near-critical Faraday modes (\ref{sec:reconstruct_mean_wave}, Theory).
%In contrast, a perturbative Floquet analysis of the system's Faraday modes demonstrates that, unlike ordered geometries with simple symmetries \cite{HMFCB2013, Saenz2018}, the contribution of Faraday modes to the mean wave field cannot be adequately characterised in terms of a small number of modes. 
The mean wave field may thus be regarded as a long-wave potential modulated by a unstructured high-frequency scatter of Faraday waves across the disordered medium.
}
% ---------------------

\vspace{4pt}
\section{Conclusions}\label{sec:conclusions}

While Anderson localization has been realized in numerous systems of classical  waves\cite{billy2008direct,roati2008Anderson,Jendrzejewski2012,dalichaouch1991microwave,wiersma1997localization,Schwartz2007Transport,segev2013anderson,Yamilov2023,hu2008localization,belzons1988gravity,devillard1988Localization},  the  original electronic localization is distinct because it incorporates the notion of a particle \reply{effectively} coming to a halt due to the landscape disorder\cite{anderson1958Absence,anderson2008anderson,Lagendijk2009Fifty,Kramer_1993,imada1998metal}. 
We have demonstrated that, owing to the resonant interaction with its self-excited wave field, walking  droplets over random topographies may exhibit \textit{dual} wave-particle localization analogous to that of subatomic particles in disordered media. 
\reply{The walker localization is observed in the high-memory regime, as the extent of the walker's wave field becomes comparable to that of the Faraday mode of the topography, and the waves persist for a relatively long period of time. The waves excited by the droplet at each impact form}  an effective wave potential that  draws the droplet towards the localization region, where it \reply{resonates with the most unstable cavity mode, causing the droplet} to execute loops at relatively high speed.
\reply{At low memory, when the walker's wave field is small and decays relatively quickly, the droplet can easily traverse the random topography, ultimately exhibiting diffusive motion.}
%Observed when the guiding wave field extends over a wide region of the random landscape at high memory,
%It is important to highlight the differences between the wave-like statistical behavior of  walking droplets and that expected for particles undergoing Fokker–Planck-type dynamics\cite{Hanggi1990,METZLER20001} in the same weak landscape. Notably, the walker's characteristic kinetic energy is largest at the localization region (Fig.\,\ref{fig:main_Fig_1}\fp{e}), suggesting a spatially-varying `temperature' distribution. The localization region  would thus correspond to the noise-activated area with the highest escape rate\cite{Hanggi1990,METZLER20001};  the region where a stochastic active particle would spend the least amount of time. 
In a broader context, the suppression of diffusion observed in the walker system as the memory increases invites the investigation of localization effects in other active systems with memory effects
\cite{Maass2017,Michelin2023} and wave-guided motion, including  those realized with stratified flows \cite{ludion}, canoes 
 \cite{gunwale}, capillary surfers \cite{surfer} and acoustically-forced bubbles \cite{acoustic}. The  dual wave-particle localization of walking droplets also motivates new directions of research in wave\cite{filoche2012universal,PhysRevLett.116.056602} and many-body\cite{RevModPhys.91.021001,choi2016exploring} localization, as well as  topological active matter\cite{hasan2010topological,meier2018observation,Shankar2022}.

\vspace{6pt}
\begin{footnotesize}
\noindent
\textbf{{Acknowledgments.}}
The authors thank Rodolfo R. Rosales for enlightening discussions.  
P.J.S. gratefully acknowledges financial support from the Alfred P. Sloan Foundation (Sloan Research Fellowship) and U.S. National Science Foundation (CAREER award CBET-2144180). 
A.J.A. gratefully acknowledges financial support from UNC's Office of Undergraduate Research (Summer Undergraduate Research Fellowship).
\textbf{Data availability.} The data that supports the findings of this study are available from the corresponding author upon reasonable request.

\end{footnotesize}

%\clearpage

%\section*{APPENDIX}

%\section{INTRODUCTION}

%We combine experiments, simulations and theory to investigate the wave-particle localization of walking droplets over submerged random topographies. We first describe, in section \ref{sec:SI_experiment}, the setup, parameter regime, and data acquisition protocol used in our experiments.  In section \ref{sec:SI_theory}, we describe the reduced hydrodynamic pilot-wave model used to capture the experimental observations, along with a perturbative Floquet analysis that rationalizes the relationship between the instantaneous wave modes, the mean pilot-wave and the particle localization when the system is in statistical equilibrium. In section \ref{sec:SI_quantum_localization}, we describe the Schr\"odinger simulations used to illustrate the quantum phenomenon of Anderson localization. In section \ref{sec:SI_simulations}, we present the simulations of the walker model that reproduce and extend the experimental observations (section \ref{sec:SI_walker_simulation}), and the simulations of the waveless-particle model that we use to contextualize the distinct walker behavior (section \ref{sec:SI_waveless_simulations}). A series of supplementary figures is included in section \ref{sec:SI_Supfigs}. Our investigation also includes three supporting videos. The corresponding legends are included in section (Supplementary Information)).

% ----------------------------------------------
%               EXPERIMENTS
% ----------------------------------------------
\renewcommand\thesection{Appendix \Alph{section}}
\renewcommand\thesubsection{\arabic{subsection}}
\renewcommand\thesubsubsection{\roman{subsubsection}}
\renewcommand{\tablename}{Appendix Table}
\renewcommand*\thetable{\arabic{table}}

%\newcounter{mycounter}
\setcounter{section}{0}

%\newrefformat{SI}{Appendix \ref{#1}}

\newcommand{\parder}[2]{\frac{\partial #1}{\partial #2}}
\newcommand{\totder}[2]{\frac{\mathrm{d} #1}{\mathrm{d} #2}}

\section{EXPERIMENTS \label{sec:SI_experiment}}

We  describe here the experimental set-up (\ref{sec:experimental_set-up}), and data acquisition protocol (\ref{sec:data_aquisition}) adopted to obtain the results presented in the main text.

\begin{figure}[h!]
\includegraphics[scale=1]{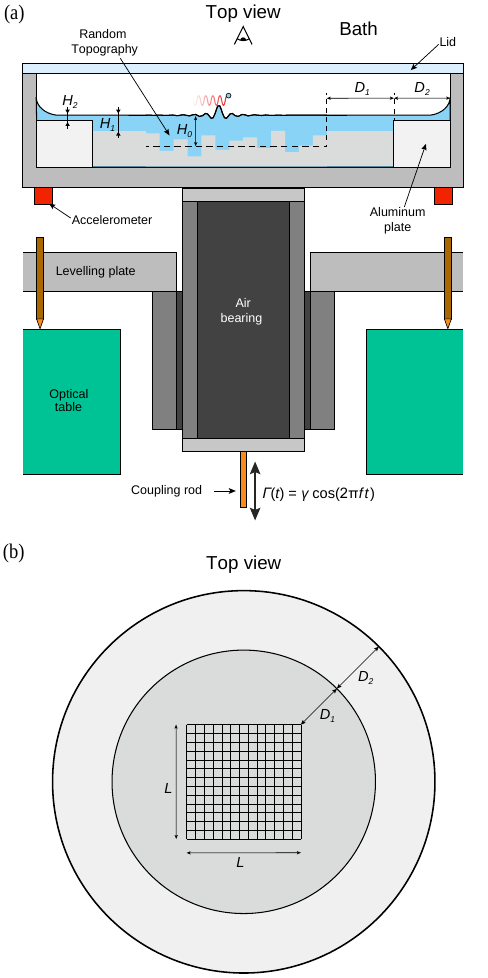}
\caption{\textbf{Schematic of the experimental set-up.} 
(a) The test section was mounted on an optical table and vibrated vertically by an electromagnetic shaker. The shaker was connected to the bath by a thin steel rod coupled with a linear air bearing to minimize  lateral vibrations. The forcing acceleration was monitored by two piezoelectric accelerometers. The bath was sealed with a transparent acrylic lid to ensure that ambient air currents did not affect the experiments.
(b) The square random topography of width $L=68.25$\,mm was surrounded by a shallow layer of depth $H_1=1.20$\,mm and minimum width $D_1=30.5$\,mm, and a thinner damping layer of depth $H_2=1.10$\,mm and  width $D_2=35.6$\,mm to prevent meniscus waves from influencing the droplet dynamics.
}
\label{fig:sup_fig_schematic}
\end{figure}

\subsection{Experimental set-up\label{sec:experimental_set-up}} 

A schematic of the experimental setup is presented in Fig.\,\ref{fig:sup_fig_schematic}. 
The random topography was composed of a regular grid of $13\times13$ square tiles, each with a width $W=5.25$\,mm (matching the Faraday wavelength, $\lambda_\mathrm{F}$) and a random height drawn from the uniform distribution in the interval $\Delta H=\pm 0.3$\,mm. The topography included $4\times4$ regularly spaced pillars with height  $\Delta H=0.6$\,mm to facilitate  data acquisition; the pillars play no significant role in the localization problem (\ref{sec:pillars}).
The variable topography was 3D printed with a FormLabs Form 3+ stereolithography printer (25\,$\mu$m XY resolution, 25\,$\mu$m minimum layer thickness, 85\,$\mu$m laser spot size, black photopolymer resin FormLabs V4) with a minimum base thickness of $0.25$\,mm and vertically oriented to minimize height deformations during the curing process. The prints were washed for 30\,min in isopropanol beforehand, and later cured for 30\,min under UV light at $60\,{}^{\circ}$C. The print supports were kept on the parts during the curing process to provide extra stability, thus minimizing warping due to thermal expansion. 

The random topography was bolted to the base of a circular aluminum bath, which was later filled with silicone oil with density $\rho=950$ kg\,m$^{-3}$, viscosity $\nu=20.9$\,cSt, and surface tension $\sigma=20.6$\,mN\,m$^{-1}$. The heterogeneous domain, with base depth $H_0=1.85$\,mm, was surrounded by a shallower liquid layer, with uniform depth $H_1=1.20$\,mm and minimum width $D_1=30.5$\,mm, which prevented the droplet from escaping the region of interest. An additional outer layer of depth $H_2=1.10$\,mm and width $D_2=35.6$\,mm was located between the 3D printed topography and the bath's wall to damp the meniscus waves\cite{Douady1990}. 
The aluminum bath was sealed with a transparent acrylic lid to ensure that external air currents did not influence the experiment.

The bath was subject to  vertical sinusoidal oscillations with frequency $f=\omega_0/2\pi=70$\,Hz using a state-of-the-art vibrating set-up\cite{Harris2015a} whose precision has enabled the characterization of the walker dynamics in numerous settings\cite{Bush2015ARFM,Bush2021Hydrodynamic}. Specifically, the aluminum bath was mounted to a  linear air bearing  (PI L.P., $4\times 4''$ cross-section, $6.5''$ long hollow bar) connected to an air-cooled electromagnetic shaker (Modal Shop, 2110E) powered by an external power amplifier (Modal Shop, 2050E09-FS). The air bearing rests on an optical table (Newport,  custom SG-34-4 breadboard, $3.0'\times 4.0'\times 4.32''$, centered $8.00''$ diameter access hole) for high-precision leveling through three micrometer screws (Newport, AJS100-1H). The shaker is connected to the inner slide bar of the air bearing through a thin coupling rod (diameter 1.6\,mm, length 60\,mm) with very low lateral stiffness to prevent any non-axial vibration generated by the shaker from being transmitted to the playload. Two  piezoelectric accelerometers (PCB, 352C65) mounted in diametrically opposed locations atop the slider bar measure the vibration amplitude, which is specified and monitored through a custom closed-feedback loop developed using a data acquisition system (NI, USB-6343) and Labview software. This set-up\cite{Harris2015a} avoids undesirable resonances and multi-directional vibrations, achieving a precision of the driving acceleration of $\pm0.002g$ and 
spatially uniform vibration within 0.1\%.

We used a piezoelectric droplet-on-demand generator\cite{Harris2015} to produce droplets of the same silicone oil and reproducible size (radius $R=0.371\pm0.003$\,mm). The bath was illuminated with a LED light ring to increase the contrast between the drops and the black background. The motion of the walking droplet was recorded at 10 frames pers second with a CCD camera (AV Mako U-130B USB3, Mono, $6.2$\,px/mm) mounted directly above the bath, and tracked with an in-house particle-tracking algorithm written in Matlab. The form of the wave field was captured by recording the normal reflection of light at the free surface\cite{Douady1990} through a semi-reflective mirror placed at $45^{\circ}$  between the CCD camera and the bath. For the waves visualization, the light ring was replaced by a square diffuse-light lamp facing the mirror horizontally. One then observes images with bright regions  corresponding to horizontal parts of the surface, extrema or saddle points\cite{Douady1990}.

% ----------------------------------------------
%              DATA ACQUISITION
% ----------------------------------------------

\subsection{Data acquisition\label{sec:data_aquisition}}

We define the Faraday threshold, $\gamma_\mathrm{F}=3.775g$, as the critical vibrational acceleration at which Faraday waves\cite{Faraday1831} are first observed above the localization region in the absence of the droplet. To ensure a steady state in the bath, the experimental set-up was left vibrating at $\gamma\approx\gamma_F$ for at least 2\,h before conducting any experiment. The droplet motion was recorded for $6$\,h in 20-min intervals. The length of the interval was selected to maintain the prescribed memory, $\gamma/\gamma_F=99.8\pm0.1\%$, which may drift slowly due  to weak variations in viscosity and surface tension resulting from ambient temperature changes. 
The Faraday threshold was measured before and after each experimental run to ensure that the Faraday threshold was never crossed. Thus, no waves existed in the bath in the absence of the droplet.
%With this protocol, the variation in each experimental run was limited to \todo{$|\Delta\gamma_\mathrm{F}|/\gamma_\mathrm{F}<??$}, where $\Delta\gamma_\mathrm{F}$ represent the variation in the Faraday threshold over a 20-m interval.
We recorded the droplet motion for 6 h in total, which corresponded
to approximately 750,000 bounces and roughly 2,300 domain
crossings for the  average speed $v_0=7.3$\,mm/s. 

At the prescribed forcing acceleration, the memory time\cite{Eddi2011} is $T_\mathrm{M}=T_\mathrm{d}/(1-\gamma/\gamma_\mathrm{F})\approx 26$\,s, where $T_\mathrm{d}=\lambda_\mathrm{F}^2/(8\pi^2\nu)\approx0.0525$\,s is the characteristic wave decay time in the absence of vibrational forcing\cite{MB2013b}. The memory time is thus negligible relative to length of the recording intervals, which guarantees that the emergent statistics were not influenced by initial conditions. We also confirmed ergodicity with our accompanying walker simulations (\ref{sec:SI_simulations}).

Faraday waves are subharmonic, oscillating at half of the prescribed vibrational frequency. The Faraday wavelength, $\lambda_\mathrm{F}=2\pi/k_\mathrm{F}$, may thus be obtained through the capillary-gravity dispersion relation\cite{KT1994},  $\omega_0^2/4 = (gk_\mathrm{F} + \sigma k_\mathrm{F}^3/\rho)\tanh(k_\mathrm{F}H)$, where $g$ is the gravitational acceleration and $H$ is the fluid depth. For the base depth, $H_0=1.85$\,mm, the  characteristic Faraday wavelength is $\lambda_\mathrm{F}=5.23$\,mm. The difference in $\lambda_\mathrm{F}$ due to the maximum depth variations ($\Delta H =\pm0.3$\,mm) and pillars ($\Delta H =+0.6$\,mm) is at most 1\% and 3\%, respectively.

We estimated the walker's spatial extent, $l$, or  `damping length'\cite{Tadrist2018a}, through direct  visualization of the experimental wave field at the prescribed memory. Our measurements, $l\approx 7\lambda_\mathrm{F}$, are in the same order as the damping length predicted by theory\cite{Tadrist2018a}.

% ----------------------------------------------
%                   THEORY
% ----------------------------------------------

\section{THEORY\label{sec:SI_theory}}

We combine simulations and analytical techniques to characterise the dynamics and statistics of the walking droplet system. We first describe a reduced theoretical model (\ref{sec:pilot-wave_model}), which forms the basis of our simulations and accompanying analysis. To distinguish the prevailing features of this pilot-wave system, we decompose the fluid evolution into the dynamics of long waves and Faraday modes (\ref{sec:linear_system}), the latter of which is approximated using Floquet theory (\ref{sec:Faraday_modes}). Finally, we characterise the manifestation of long waves and Faraday modes in the formation of the time-averaged pilot-wave arising when the system is in statistical equilibrium (\ref{sec:mean_wave_field}).

% ----------------------------------------------
%               PILOT-WAVE MODEL
% ----------------------------------------------
\subsection{Pilot-wave model}
\label{sec:pilot-wave_model}

We present a reduced theoretical model\cite{Faria2016} of the pilot-wave system evolving over variable topography, which we utilise to buttress experimental observations and shed light on the dynamical and statistical features of the pilot-wave system.

% ----------------------------------------------
%               FLUID EVOLUTION
% ----------------------------------------------

\subsubsection{Fluid evolution}

The fluid evolution is governed by the quasi-potential flow model\cite{MGNB2015, DDZ2008} for the free-surface elevation, $\eta(\x,t)$, and the linearized free-surface velocity potential, $u(\x,t) = \phi(\x,0,t)$, where $\phi(\x, z, t)$ satisfies Laplace's equation in the fluid bulk, with no-flux boundary conditions on the submerged topography. In the vibrating frame of reference, $u$ and $\eta$ evolve according to
\begin{subequations}
\label{eq:QPE}
\begin{align}
u_t &= -G(t)\eta + \frac{\sigma}{\rho} \Delta \eta + 2\nu \Delta u - \frac{P}{\rho}, \label{eq:QPE_u} \\
\eta_t &= \DtN u + 2\nu \Delta \eta, \label{eq:QPE_eta}
\end{align}
\end{subequations}
where $\sigma$ is the coefficient of surface tension, $\rho$ is the fluid density, $\nu$ is the fluid kinematic viscosity, $P$ is the pressure exerted by the droplet on the free surface (defined in section \ref{sec:wave_droplet_coupling}), and $\DtN$ is the Dirichlet-to-Neumann operator, defined $\DtN\phi|_{z = 0} = \phi_z|_{z = 0}$. As a consequence of expressing the fluid evolution in the vibrating frame of reference, we define $G(t) = g - \gamma \cos(\omega_0 t)$ as the effective gravitational acceleration, where  $\omega_0$ is the angular frequency,  $g$ is the gravitational acceleration in the absence of vibrational forcing, and $\gamma$ is the peak vibrational acceleration. 

We adopt the formulation of Faria\cite{Faria2016}, who, informed by shallow-water theory, approximated the Dirichlet-to-Neumann operator for variable topography by the phenomenological operator $\DtN = -\nabla \cdot b(\x)\nabla$, where the effective depth, $b(\x)$, is defined in terms of the actual fluid depth, $H(\x)$. Specifically, 
$$b(\x) = \frac{\tanh(k_c(\x)H(\x))}{k_c(\x)} $$
and $k_c = \mathrm{argmin}_k \gamma_c(k)$ is the critical wavenumber for each value of $\x$, where\cite{MGNB2015}
$$\gamma_c^2(k)= \frac{4g^2}{\omega_g^4(k)}\bigg(\bigg[\omega^2(k) + \mu^2(k) - \frac{\omega_0^2}{4}\bigg]^2 + \omega_0^2\mu(k)^2\bigg) $$
determines the neutral stability curve,
$\omega^2(k) = k\tanh(kH)(g + \sigma k^2/\rho)$ is the finite-depth  gravity-capillary dispersion relation, $\mu(k) = 2\nu k^2$ determines wave damping, and $\omega_g^2(k) = gk \tanh(kH)$ is the finite-depth dispersion relation for gravity waves.

% ----------------------------------------------
%            WAVE-DROPLET COUPLING
% ----------------------------------------------
\subsubsection{Wave-droplet coupling}
\label{sec:wave_droplet_coupling}

We model the wave-droplet coupling by prescribing instantaneous periodic impacts\cite{Faria2016, DM2017} at times $t_n = nT_\mathrm{F} + \theta_I/\omega_0$ (for $n = 0,1,2,\ldots$), where $T_\mathrm{F} = 4\pi/\omega_0$ is the Faraday period and $\theta_I$ is the impact phase. We thus prescribe an interaction force of the form 
\begin{equation}
\label{eq:force}
F(t) = mg T_\mathrm{F} \sum_{n = 0}^\infty\delta(t - t_n),
\end{equation}
where $m=\frac{4}{3}\pi\rho R^3$ is the droplet mass, $R$ is the droplet radius, and $\delta$ is the Dirac delta function\cite{Faria2016, DM2017}. Moreover, we prescribe that the pressure, $P$, exerted on the free surface during each impact is localized to a point in space about the droplet's current position, $\x_\mathrm{p}(t)$, namely $P(\x, t) = F(t) \delta (\x - \x_\mathrm{p}(t))$\cite{Faria2016, DM2017}.
We model the droplet's horizontal motion according to\cite{MB2013b} 
\begin{equation}
\label{eq:drop_trajectory}
m \ddot{\x}_\mathrm{p} + D(t) \dot{\x}_\mathrm{p} = - F(t) \nabla \eta(\x_\mathrm{p}, t),
\end{equation}
where
\[ D(t) = c_4 \sqrt{\frac{\rho R}{\sigma}} F(t) + 6 \pi R \mu_{\text{air}} \]
is the drag acting on the droplet,
 $\mu_{\text{air}
}$ is the dynamic viscosity of air,  and $c_4$ is the coefficient of tangential restitution. During droplet flight, i.e.\ $t_n < t < t_{n+1}$, the coupling force vanishes, i.e.\ $F(t) = 0$, and the droplet inertia is balanced by Stokes drag. Across each impact, we apply the jump condition in the droplet's velocity, 
\begin{equation}
\label{eq:drop_trajectory_delta_jump}
[\dot\x_\mathrm{p}(t_n)]_-^+ = -\big(1 - \me^{-C}\big)\bigg(\frac{g T_F}{C}\nabla \eta(\x_\mathrm{p}(t_n), t_n) + \dot\x_\mathrm{p}(t_n^-)\bigg), 
\end{equation}
where $C = c_4 gT_\mathrm{F}\sqrt{\rho R/\sigma}$\cite{Faria2016, DM2017}.

% ----------------------------------------------
%               LINEAR SYSTEM
% ----------------------------------------------

\subsection{Wave field evolution}
\label{sec:linear_system}

We start by transforming equation \eqref{eq:QPE} into Fourier space, for which we find that the conjugate variables $\hat{u}(\kb, t)$ and $\hat{\eta}(\kb, t)$ evolve according to
\begin{subequations}
\label{eq:QPE_FT}
\begin{align}
\hat u_t &= -G(t)\hat \eta - \frac{\sigma}{\rho} |\kb|^2 \hat \eta - 2\nu |\kb|^2 \hat u - \frac{F(t)}{\rho}\me^{-\mathrm{i} \kb \cdot \x_\mathrm{p} }, \label{eq:QPE_FT_u} \\
\hat \eta_t &= \hat {\DtN_{\kb}} \hat u - 2\nu |\kb|^2 \hat \eta, \label{eq:QPE_FT_eta}
\end{align}
\end{subequations}
where $\kb$ is the wave vector and
$\hat {\DtN_{\kb}}$ is the Fourier symbol of the operator $\DtN$. For Faria's model, $ \hat {\DtN_{\kb}} \hat{u} = \kb \cdot \FT \big[b(\x) \FT^{-1} [\kb \hat{u}]\big]$, where $\FT$ and $\FT^{-1}$ represent the forward and inverse Fourier transforms, respectively.\cite{Faria2016}
We consider the fluid evolution on a doubly-periodic domain using the discrete Fourier Transform, giving rise to the nonzero discrete wave vectors $\kb_j$ for $j = 1,\ldots,\mathscr{N}$. We exclude the zero wavevector from this set as conservation of mass implies that $\hat{\eta}(\bm{0}, t) = 0$ and additive invariance of the velocity potential means that we may choose $\hat{u}(\bm{0}, t) = 0$. We proceed now to reformulate the system \eqref{eq:QPE_FT} as a matrix-vector system in Floquet form.

By defining row vectors $\hat{\bm{u}}(t)$ and $\hat{\bm{\eta}}(t)$ with elements $\hat{u}(\bm{k}_j, t)$ and $\hat{\eta}(\bm{k}_j, t)$, respectively, we may thus express \eqref{eq:QPE_FT} in matrix-vector form
\begin{subequations}
\label{QPE_FT_matrix}
\begin{equation}
\label{QPE_FT_matrix_syst}
\totder{\vb}{t} = \Big(\Ab + \gamma \cos(\omega_0 t) \Bb\Big)\vb - \bm{F}(t), 
\end{equation}
where $\vb = (\hat{\bm{u}}, \hat{\bm{\eta}})^T$,
\begin{equation}
\label{QPE_FT_matrix_def}
\Ab = \begin{pmatrix}
-2\nu\bm{K}^2 & -\big(g\bm{I} + \sigma \rho^{-1} \bm{K}^2\big) \\ 
\bm{L} & - 2\nu \bm{K}^2,
\end{pmatrix}
\end{equation}
$\bm{K}$ is a diagonal matrix with elements $|\kb_j|$, $\bm{I}$ is the identity matrix, $\bm{L}$ is a matrix representing the Fourier symbol of the operator $\DtN$, and 
\begin{equation}
\label{QPE_FT_matrix_def2}
\Bb = \begin{pmatrix}
\bm{0} & \bm{I} \\ \bm{0} & \bm{0}
\end{pmatrix}.
\end{equation}
\end{subequations}
Finally, the forcing term, $\bm{F}(t)$, in \eqref{QPE_FT_matrix_syst} represents the Fourier transform of the pressure, namely
$$\bm{F}(t) = \frac{F(t)}{\rho}\begin{pmatrix}
\bm{e}(\x_\mathrm{p}(t)) \\ \bm{0} \end{pmatrix}, $$
where $\bm{e}(\x)$ is a column vector (of length $\mathscr{N}$) with elements $\me^{- \mathrm{i} \kb_j\cdot \x}$.
Between successive impacts, the coupling force vanishes, i.e.\ $\bm{F}(t) = \bm{0}$. By assuming that $\x_\mathrm{p}(t)$ is continuous across each impact, we use equation \eqref{QPE_FT_matrix_syst} to derive the jump condition $[\vb(t_n)]_-^+ = -\bm{f}_n$\cite{Faria2016, DM2017}, where $\bm{f}_n = \bm{f}(\x_\mathrm{p}(t_n))$ and
\begin{equation}
\label{eq:f(x) forcing}
\bm{f}(\x) = \frac{mg T_\mathrm{F} }{\rho}\begin{pmatrix}
    \bm{e}(\x) \\ \bm{0}
\end{pmatrix}.
\end{equation}
Our investigation is focused, henceforth, on the computational analysis of \eqref{QPE_FT_matrix}, which governs the evolution of the $2\mathscr{N}$ complex variables describing the fluid state.

% ----------------------------------------------
%               DECOMPOSITION
% ----------------------------------------------

\subsubsection{Wave field decomposition}
\label{sec:decomposition}

In the absence of vibrational forcing, the quasi-potential flow model \eqref{eq:QPE} exhibits a viscous damping rate of $2\nu|\kb|^2$ for each Fourier mode, corresponding to the relatively slow decay of long traveling waves. When the vibrational acceleration is instead just below that of the Faraday threshold, the slow decay of long waves ($|\kb| \ll k_\mathrm{F}$) persists, and the decay time of the Faraday modes ($|\kb| \approx k_\mathrm{F}$) increases, dilating with increased proximity to the Faraday threshold (see \ref{sec:Faraday_modes}). The wave field generated by a walking droplet is thus composed of both long traveling waves and shorter standing waves (Faraday modes) \cite{MGNB2015}. We proceed to exploit the separation of length scale exhibited by the long waves and  Faraday modes to decompose the fluid evolution into a long-wave model and an approximation for near-critical Faraday modes. Specifically, we consider $\vb \approx \vb_\mathrm{L} + \vb_\mathrm{F}$, where we determine an approximate evolution equation for the long waves, denoted $\vb_\mathrm{L}$, in section \ref{sec:long_waves}, and determine an approximate solution for the near-critical Faraday modes, denoted $\vb_\mathrm{F}$, in section \ref{sec:Faraday_modes}.
We then combine our analytical results to characterise the manifestation of long waves and Faraday waves in both the instantaneous and time-averaged pilot wave.

% ----------------------------------------------
%               LONG WAVES
% ----------------------------------------------

\subsubsection{Evolution of long waves}
\label{sec:long_waves}

The submerged topography is shallow relative to the long-wave modes, and so the critical vibrational forcing for long waves ($|\kb| \ll k_\mathrm{F}$) vastly exceeds that of the Faraday threshold. \reply{Specifically, the magnitude of their critical vibrational threshold has size $O(k^{-2})$ as $k\rightarrow 0$\cite{Kumar1996}.} Hence, when the vibrational forcing is just below the Faraday threshold, the parametric forcing only has a  weak effect on the evolution of long waves. A leading-order approximation for long waves is thus afforded by setting $\gamma = 0$ in the quasi-potential flow equations \eqref{eq:QPE}, indicating that $\bm{v}_\mathrm{L}$ evolves according to
\begin{equation}
\label{eq:QPE_system_forcing_long}
\totder{\vb_\mathrm{L}}{t} = \Ab\vb_\mathrm{L} - \bm{F}(t),
\end{equation}
where the forcing, $\bm{F}(t)$, vanishes during flight ($t_n < t < t_{n+1}$) and gives rise to the jump condition $[\vb_\mathrm{L}(t_n)]_-^+ = -\bm{f}_n$ across each impact. We note that waves that are not long are strongly damped when $\gamma = 0$, owing to the dissipation rate $2\nu|\kb|^2$, and so have a negligible role in the resulting dynamics. %Although one might further reduce the long-wave model by neglecting the effects of surface tension, which is weak relative to gravity for long waves, we choose to retain this term in our formulation.

% ----------------------------------------------
%           COMPUTING FARADAY MODES
% ----------------------------------------------
\subsection{Evolution of Faraday modes}
\label{sec:Faraday_modes}

We proceed to seek approximate solutions for near-critical Faraday modes satisfying
\begin{equation}
\label{eq:QPE_system_forcing_Faraday} 
\totder{\vb_\mathrm{F}}{t} = \Big(\Ab + \gamma \cos(\omega_0 t) \Bb\Big)\vb_\mathrm{F} - \bm{F}(t),
\end{equation}
where the forcing, $\bm{F}(t)$, vanishes during flight ($t_n < t < t_{n+1}$) and gives rise to the jump condition $[\vb_\mathrm{F}(t_n)]_-^+ = -\bm{f}_n$ across each impact. Owing to the periodic parametric forcing, $\gamma \cos(\omega_0 t)\bm{B}$, we observe that equation \eqref{eq:QPE_system_forcing_Faraday} is of Floquet form during droplet flight. We thus use Floquet theory to approximate the evolution of the near-critical Faraday modes during droplet flight, giving rise to a reduced framework for characterising the manifestation of Faraday modes in the instantaneous and time-averaged pilot wave. We formulate the Floquet analysis in section \ref{sec:Floquet_theory}, before determining the critical vibration of each Faraday mode (section \ref{sec:critical_forcing}) and the corresponding near-critical decay rate (section \ref{sec:decay_rates}).

\subsubsection{Approximating Faraday modes}
\label{sec:Floquet_theory}

For any fixed value of $\gamma \neq 0$, we deduce from Floquet theory that $\vb_\mathrm{F}(t)$ has linearly independent solutions
\begin{equation}
\label{eq:vb_Floquet}
\vb_{\mathrm{F},j}(t) = \pb_j(t)\me^{\lambda_j t} \quad\mathrm{for}\quad j = 1,\ldots,2\mathscr{N},
\end{equation}
where $\pb_j(t)$ is a periodic function of period $T_\mathrm{F}$ and $\lambda_j(\gamma)$ is a Floquet exponent. We may represent $\pb_j(t)$ as a Fourier series and use as many terms are necessary in the truncated series expansion for the predicted Faraday threshold (and associated mode) to converge\cite{KT1994, Kumar1996}. However, we find that a favourable, and computationally tractable, approximation of the Faraday modes is afforded by considering only the leading-order frequencies in the Fourier series of $\pb_j(t)$, an assumption that has been utilized throughout the walking-droplet literature\cite{MB2013b, MGNB2015}.

We proceed by writing
\begin{equation}
\label{eq:p_def}
\pb_j(t) = \x_j \sin\bigg(\frac{\omega_0 t}{2}\bigg) + \yb_j \cos\bigg(\frac{\omega_0 t}{2}\bigg),
\end{equation}
where $\x_j(\gamma)$ and $\yb_j(\gamma)$ are as-yet-undetermined nonzero vectors, and then we substitute equations \eqref{eq:vb_Floquet}--\eqref{eq:p_def} into \eqref{eq:QPE_system_forcing_Faraday} and simplify the resultant expressions. By using the identities 
\begin{align*}
    2\cos(2\theta)\cos\theta &= \cos(3\theta) + \cos\theta, \\
    2\cos(2\theta)\sin\theta &= \sin(3\theta) - \sin\theta,
\end{align*}
we obtain the system of equations
\begin{multline*}
\bigg[-\frac{\omega_0}{2} \yb_j + \lambda_j \x_j - \Ab \x_j + \frac{\gamma}{2}\Bb \x_j\bigg]\sin\bigg(\frac{\omega_0 t}{2}\bigg)\me^{\lambda_j t}  \\
+ \bigg[\frac{\omega_0}{2} \x_j + \lambda_j \yb_j - \Ab \yb_j - \frac{\gamma}{2}\Bb \yb_j\bigg]\cos\bigg(\frac{\omega_0 t}{2}\bigg)\me^{\lambda_j t} \\ + \mbox{higher harmonics} = 0.
\end{multline*}
By neglecting the higher harmonics, \reply{consistent with our leading-order frequency approximation,} and utilizing the linear independence of the functions $\sin\big(\frac{1}{2}\omega_0 t\big)\me^{\lambda_j t}$ and $\cos\big(\frac{1}{2}\omega_0 t\big)\me^{\lambda_j t}$, we find that $\x_j$ and $\yb_j$ satisfy
\begin{subequations}
\label{eq:c1_c2_syst}
\begin{align}
-\frac{\omega_0}{2} \yb_j + \lambda_j \x_j - \Ab \x_j + \frac{\gamma}{2}\Bb \x_j = \bm{0}, \label{eq:c1_c2_syst1} \\
\frac{\omega_0}{2} \x_j + \lambda_j \yb_j - \Ab \yb_j - \frac{\gamma}{2}\Bb \yb_j = \bm{0}. \label{eq:c1_c2_syst2}
\end{align}
\end{subequations}
To further simplify, we eliminate $\yb_j$ by pre-multiplying \eqref{eq:c1_c2_syst1} by $[\lambda_j\bm{I} - \Ab - (\gamma/2)\Bb]$ and utilizing \eqref{eq:c1_c2_syst2}; we thus arrive at the condition $\bm{M}(\lambda_j, \gamma)\x_j = \bm{0}$, where 
\begin{equation}
\label{eq:Mdef}
\bm{M}(\lambda, \gamma) = \frac{\omega_0^2}{4}\bm{I} + \bigg[\bm{A} - \lambda\bm{I} + \frac{\gamma}{2}\bm{B}\bigg] \bigg[\bm{A} - \lambda\bm{I} - \frac{\gamma}{2}\bm{B}\bigg].
\end{equation}
The vector $\yb_j$ may then be recovered from equation \eqref{eq:c1_c2_syst1},
%, namely
%$$\bm{y}_j = \bigg(\frac{2(\lambda_j \bm{I} - %\bm{A}) + \Gamma\bm{B}}{\omega_0}\bigg) \x_j, $$
from which one may then determine the sinusoidal oscillation of the fluid, $\pb_j(t)$.

For given $\gamma \neq 0$, we seek a series of nonzero vectors, $\x_j(\gamma)$, and growth rates, $\lambda_j(\gamma)$, defined by the nonlinear eigenvalue problem $\bm{M}(\lambda_j, \gamma)\x_j = \bm{0}$. Then, we define the critical vibrational forcing, $\gamma_j$, for each mode, with $\lambda_j(\gamma_j) = 0$, and denote $\x_{j,0} = \x_j(\gamma_j)$. We now develop methodologies for (i) determining $\gamma_j$ and $\x_{j,0}$ (section \ref{sec:critical_forcing}) and (ii) determining $\lambda_j(\gamma)$ and $\x_j(\gamma)$ for $|\gamma - \gamma_j|/g \ll 1$ (section \ref{sec:decay_rates}). These results are then applied to the computation of the mean pilot wave in section \ref{sec:mean_wave_field}.

% ----------------------------------------------
%      INSTABILITY THRESHOLD FOR EACH MODE
% ----------------------------------------------

\subsubsection{Critical forcing for each mode}
\label{sec:critical_forcing}

The critical vibrational forcing, $\gamma_j$, for each eigenmode, denoted $\x_{j,0} \neq \bm{0}$, arises when the corresponding decay rate vanishes, i.e.\ $\lambda_j = 0$, with $\bm{M}(0, \gamma_j)\x_{j,0} = \bm{0}$. In order to determine $\gamma_j$ and $\x_{j,0}$, we rearrange equation \eqref{eq:Mdef} and note that $\bm{B}^2 = \bm{0}$, giving 
$$\bm{M}(0, \gamma) = \frac{\omega_0^2}{4}\bm{I} + \Ab^2 + \frac{\gamma}{2}\Big[\Bb\Ab - \Ab\Bb\Big]. $$
From $\bm{M}(0, \gamma_j)\x_{j,0} = \bm{0}$, we deduce that
$$\bigg(\bm{A}^2 + \frac{\omega_0^2}{4}\bm{I} \bigg) \x_{j,0} = \frac{\gamma_j}{2}\Big[\Ab\Bb - \Bb\Ab\Big]\x_{j,0}. $$
Finally, by writing 
\[\mathcal{A} = \bm{A}^2 + \frac{\omega_0^2}{4}\bm{I} \quad\mathrm{and}\quad \mathcal{B} = \frac{1}{2}\Big[\Ab\Bb - \Bb\Ab\Big],\]
we observe that $\bm{M}(0, \gamma_j)\x_{j,0} = \bm{0}$ reduces to the generalized eigenvalue problem $\mathcal{A}\x_{j,0} = \gamma_j \mathcal{B}\x_{j,0}$, where $\gamma_j$ is an eigenvalue (indexed by $j$) and $\x_{j,0}$ is a corresponding eigenvector. 
It is expedient, however, to reformulate the generalized eigenvalue problem as\cite{KT1994, Kumar1996}
\begin{equation}
\label{eq:eig_prob}
\mathcal{A}^{-1}\mathcal{B}\x_{j,0} = \frac{1}{\gamma_j}\x_{j,0},
\end{equation}
from which it follows that $\gamma_j$ is the reciprocal of an eigenvalue of the matrix $\mathcal{A}^{-1}\mathcal{B}$.

Notably, choosing to instead eliminate $\bm{x}_j$ from system \eqref{eq:c1_c2_syst} gives rise to the similar relationship 
$$\mathcal{A}^{-1}\mathcal{B}\bm{y}_{j,0} = -\frac{1}{\gamma_j}\bm{y}_{j,0}. $$
We thus deduce that if $\alpha$ is an eigenvalue of the matrix $\mathcal{A}^{-1}\mathcal{B}$, then so is $-\alpha$. Consequently, the critical thresholds appear in negated pairs, consistent with the invariance of the fluid system \eqref{eq:QPE} under the mapping $(t,\gamma) \mapsto (t + \pi/\omega_0, -\gamma)$. Modes with $\gamma_j > 0$ are damped for $\gamma < \gamma_j$ and excited for $\gamma > \gamma_j$; in contrast, modes with $\gamma_j < 0$ are excited for $\gamma < \gamma_j$ and damped for $\gamma > \gamma_j$. When considering $\gamma > 0$, we conclude that modes satisfying $\gamma_j < 0$ are strongly damped.

% ----------------------------------------------
%           NEAR-CRITICAL DECAY RATES
% ----------------------------------------------

\subsubsection{Mode decay rates}
\label{sec:decay_rates}

For the remainder of this section, we restrict our attention to the case $\gamma > 0$, and define the Faraday threshold $\gamma_\mathrm{F} = \min_j |\gamma_j|$. For a given value of $\gamma < \gamma_\mathrm{F}$, we seek to determine the decay rate, $\lambda_j(\gamma) < 0$, of each wave mode. Specifically, we must find $\lambda_j$ and $\x_j(\gamma) \neq 0$ so that $\bm{M}(\lambda_j, \gamma)\x_j = \bm{0}$. Such a problem is computationally challenging owing to the nonlinear appearance of $\lambda_j$ in equation \eqref{eq:Mdef}. Instead, we focus on the relevant and asymptotically tractable case of near-critical forcing, for which the vibrational acceleration, $\gamma$, is close to the critical vibrational forcing, $\gamma_j$, for the given eigenmode. For this section, we ignore modes satisfying $\gamma_j < 0$, which are strongly damped (and so fall outside of the regime of near-critical forcing).

We proceed by seeking an asymptotic expansion in terms of the parameter $\epsilon_j = (\gamma_j - \gamma)/g$, where $0 < \epsilon_j \ll 1$ for near-critical modes. Specifically, we posit
\begin{subequations}
\label{eq:asymp_exp}  
\begin{align}
    \lambda_j &\sim \epsilon_j \lambda_{j,1} + O(\epsilon_j^2), \\
    \x_j &\sim \x_{j,0} + \epsilon_j \x_{j,1} + O(\epsilon_j^2), 
\end{align}
\end{subequations}
where $\x_{j,0} \neq \bm{0}$ satisfies $\bm{M}(0, \gamma_j)\x_{j,0} = \bm{0}.$
Upon substituting the asymptotic expansions \eqref{eq:asymp_exp} and perturbation $\gamma = \gamma_j - g\epsilon_j$ into the equation $\bm{M}(\lambda_j, \gamma)\x_j = \bm{0}$, we Taylor expand the arguments of $\bm{M}$. After some algebra, we obtain
\begin{multline*}
    \lambda_{j,1}\bm{M}_\lambda(0, \gamma_j)\x_{j,0} - g\bm{M}_\gamma(0, \gamma_j)\x_{j,0} \\ + \bm{M}(0,\gamma_j)\x_{j,1} = O(\epsilon_j),
\end{multline*}
where subscripts denote partial differentiation of the matrix $\bm{M}(\lambda, \gamma)$. By neglecting terms of size $O(\epsilon_j)$, we observe that $\x_{j,1}$ satisfies the linear system
\begin{equation}
    \label{eq:x_n1}
    \bm{M}(0,\gamma_j)\x_{j,1} = \big[g\bm{M}_\gamma(0, \gamma_j) - \lambda_{j,1}\bm{M}_\lambda(0, \gamma_j)\big] \x_{j,0},
\end{equation}
where we recall that $\bm{M}(0, \gamma_j)$ is a singular matrix. By the Fredholm Alternative Theorem, there exists a solution for $\x_{j,1}$ when 
\begin{equation}
    \label{eq:FAT}
    \bm{z}_{j,0}^*\big[g\bm{M}_\gamma(0, \gamma_j) - \lambda_{j,1}\bm{M}_\lambda(0, \gamma_j)\big]\x_{j,0} = 0,
\end{equation}
where $\bm{z}_{j,0}$ is a left eigenvector of the singular matrix $\bm{M}(0,\gamma_j)$. Specifically, $\bm{z}_{j,0}$ satisfies  $\bm{z}_{j,0}^* \bm{M}(0,\gamma_j) = \bm{0}$, where $*$ denotes the conjugate transpose. By rearranging equation \eqref{eq:FAT}, we thus deduce that the leading-order decay rate, $\lambda_{j,1} < 0$, is defined by the quotient
$$\lambda_{j,1} = \frac{g\bm{z}_{j,0}^*\bm{M}_\gamma(0, \gamma_j)\x_{j,0}}{\bm{z}_{j,0}^*\bm{M}_\lambda(0, \gamma_j)\x_{j,0}},$$
where $ \bm{M}_\gamma(0, \gamma_j) = \frac{1}{2}(\bm{B}\bm{A} - \bm{A}\bm{B}) $ and $\bm{M}_\lambda(0, \gamma_j) = -2\bm{A}$. Although one may, in principle, proceed to solve equation \eqref{eq:x_n1} for the higher-order correction to the eigenmode, denoted $\x_{j,1}$, we find that the leading-order form, $\x_j = \x_{j,0} + O(\epsilon_j)$, is satisfactory for our purposes. The growth rate may then be approximated by $\lambda_j \sim \lambda_{j,1}(\gamma_j - \gamma)/g$.

Finally, it remains to efficiently compute the left eigenvector, $\bm{z}_{j,0}$. Using a similar formulation to section \ref{sec:critical_forcing}, we deduce that $\bm{z}_{j,0}$ and $\gamma_j$ satisfy the generalized left-eigenvector problem
$$\bm{z}_{j,0}^*\mathcal{A} = \gamma_j\bm{z}_{j,0}^*\mathcal{B}. $$
By post-multiplying by $\mathcal{A}^{-1}$ and taking the conjugate transpose of both sides, we deduce that $\bm{z}_{j,0}$ is a right eigenvector of the matrix $(\mathcal{B}\mathcal{A}^{-1})^*$ with real eigenvalue $\gamma_j^{-1}$.

\begin{figure*}
\includegraphics[scale=1]{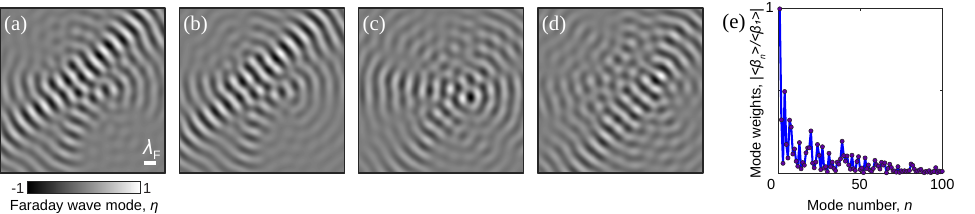}
\caption{\textbf{Theoretical Faraday eigenmodes.} Subcritcal Faraday modes with the slowest decay rates computed theoretically (section \ref{sec:Faraday_modes}). The critical thresholds are $\gamma/g = $ (a) 3.7648, (b) 3.7658, (c) 3.7828, (d) 3.7852, respectively. For $\gamma/\gamma_\mathrm{F} = 99.89\%$, the corresponding decay rates are $\lambda T_\mathrm{F} = $ (a) $-0.0016$, (b) $-0.0020$, (c) $-0.0084$, (d) $-0.0093$.
Each eigenmode is normalized by its maximum value.
(e) The absolute value of the mode weights, $|\langle \beta_n \rangle|$, normalised by that of the dominant mode,  $|\langle \beta_1 \rangle|$, for the modal composition of the mean wave field accompanying the probability density function in Fig.\ \ref{fig:sup_fig_waveless_experiment_simulations}\fp{c}. The modes are ordered with decreasing proximity to the Faraday threshold, and their weightings decrease slowly (and non-uniformly) with the mode number, $n$ \reply{(\ref{sec:Faraday_mean_wave}, equation \eqref{eq:iterative_map_average_v2})}.
%\todo{Please can you fix typo in the vertical axis label in E; the final 1 should be a subscript. Also, there should be absolute values, i.e.\ $|\langle \beta_n \rangle / \langle \beta_1 \rangle|$. Thanks.}
%\todo{Pedro's question for Matt: (1) Could you please finish/polish the caption. (2) Are the units correct? We are don't everything dimensional and I am not sure what you sent to Abel and what Abel did there. (3) If you provide the data, we can add an additional panel to show the slow decay rates. (4) Can you please add a few lines explanning somewhere in the SI?}%\textcolor{blue}{The patterns are eigenmodes, and so can be normalised in any way. I suggest normalising so that the maximum value of each panel is equal to one. If we want decay rate, then we should choose a value of $\gamma$. For now, I've just written the critical vibrational acceleration for each mode.}}
\label{fig:sup_fig_faraday_modes}}
\end{figure*}

The first four Faraday modes (ordered with increasing $\gamma_j > 0$) are presented in Fig.\,\ref{fig:sup_fig_faraday_modes}. The first mode is the dominant Faraday mode, and has the slowest decay rate for vibrational forcing just below the Faraday threshold. Notably, the predicted Faraday threshold, $\gamma_\mathrm{F} = 3.7648g$, is close to the estimation from simulations, $\gamma_\mathrm{F} = 3.806g$ (section \ref{sec:faraday_threshold}). The modes typically align across the diagonal of the cavity, with the direction selected by the relative wave excitation within the localization region. The first two modes are closely related by a phase shift, and have very similar critical thresholds and decay rates.

% ----------------------------------------------
%   EVOLUTION AND STATISTICS OF THE PILOT WAVE
% ----------------------------------------------

\subsection{Mean wave field}
\label{sec:mean_wave_field}

We proceed to consider the evolution of the pilot wave generated by successive droplet impacts, with a particular focus on the statistical convergence of the wave field according to the long-time average\cite{DMB2018, DMW2020}
$$\langle \eta\rangle (\x) = \lim_{N\rightarrow\infty}\frac{1}{N}\sum_{n = 0}^{N-1} \eta(\x, t_n). $$
Specifically, we use Floquet theory to determine an iterative map\cite{DM2017} (section \ref{sec:Faraday_map}) for the evolution of the near-critical Faraday modes, $\vb_\mathrm{F}(t)$, which we then use to derive a formula for the contribution of Faraday modes to the time-averaged wave field (section \ref{sec:Faraday_mean_wave}). We then determine the analogous contribution of long waves, $\vb_\mathrm{L}(t)$, to the time-averaged wave field (section \ref{sec:long_mean_wave}), which we couple with the contribution of Faraday modes to determine the predominant features of the mean wave field, $\langle \eta\rangle(\x)$ (section \ref{sec:reconstruct_mean_wave}).

% ----------------------------------------------
%                   FARADAY MAP
% ----------------------------------------------

\subsubsection{Faraday modes: iterative map}
\label{sec:Faraday_map}

Between successive impacts, we use the linearly independent Floquet solutions (see equation \eqref{eq:vb_Floquet}) as a basis for $\vb_\mathrm{F}(t)$. During droplet flight, i.e.\ $t_n < t < t_{n+1}$, we may thus write
$$\vb_\mathrm{F}(t) = \sum_j c_j \pb_j(t) \me^{\lambda_j t}, $$
where the $c_j$ are arbitrary constants to be determined by the initial conditions, and the sum is over all wave modes ($j = 1, \ldots, 2\mathscr{N}$). However, as we are primarily interested in the wave field at each impact (across which the free surface is continuous), we determine an iterative map for the value of the vector $\vb_\mathrm{F}(t_n^+)$, which represents the fluid variables just after each impact\cite{DM2017, DMW2020}.

To determine the iterative map, we start by expressing
\begin{equation}
\label{eq:v_map_def}
\vb_\mathrm{F}(t_n^+) = \sum_j \beta_{j,n} \pb_j(t_n) \quad\mbox{for all}\,\,\, n,
\end{equation}
where we proceed to determine the evolution of the coefficients, $\beta_{j,n}$, between successive impacts.
By exploiting the periodicity of $\pb_j(t)$, we compute
$$\vb_\mathrm{F}(t_{n+1}^-) = \sum_j \beta_{j,n} \me^{\lambda_j T_\mathrm{F}} \pb_j(t_n), $$
which we couple with the jump condition $[\vb_\mathrm{F}(t_{n+1})]_-^+ = -\bm{f}_{n+1}$ to obtain
\begin{equation}
\label{eq:v_map_next_impact}
\vb_\mathrm{F}(t_{n+1}^+) = \sum_j \beta_{j,n} \me^{\lambda_j T_\mathrm{F}} \pb_j(t_n) - \bm{f}_{n+1}.
\end{equation}
To simplify notation, we define the matrix of periodic solutions $\bm{P} = (\pb_1(t_n), \pb_2(t_n),\ldots)$, the diagonal matrix $\bm{\Lambda}$ with diagonal elements $\me^{\lambda_j T_\mathrm{F}}$ (the Floquet multipliers), and the column vector $\bm{\beta}_n$ with elements $\beta_{j,n}$. By noting that equations \eqref{eq:v_map_def} (with $n$ replaced by $n+1$) and \eqref{eq:v_map_next_impact} may be written as $\vb_\mathrm{F}(t_{n+1}^+) = \bm{P}\bm{\beta}_{n+1}$ and $\vb_\mathrm{F}(t_{n+1}^+) = \bm{P}\bm{\Lambda}\bm{\beta}_n - \bm{f}_{n+1}$, respectively, we eliminate $\vb_\mathrm{F}(t_{n+1}^+)$ to deduce the iterative map
\begin{equation}
\label{eq:iterative_map_v1}
\bm{P}\bm{\beta}_{n+1} = \bm{P}\bm{\Lambda}\bm{\beta}_n - \bm{f}_{n+1}.
\end{equation}
Finally, we use the linear independence of the periodic solutions $\bm{p}_j(t)$ to invert the matrix $\bm{P}$ in equation \eqref{eq:iterative_map_v1}, giving rise to the diagonal system
\begin{equation}
\label{eq:iterative_map_v2}
\bm{\beta}_{n+1} = \bm{\Lambda}\bm{\beta}_n - \bm{P}^{-1}\bm{f}_{n+1}.
\end{equation}
The matrix $\bm{\Lambda}$ encodes the damping of each wave mode between successive droplet impacts, and may be regarded as the system's diagonalized fundamental matrix over one Faraday period\cite{DM2017, DMW2020}. The iterative map \eqref{eq:iterative_map_v2} may be used as for simulating and analysing the dynamics of the pilot-wave system; in particular, we proceed to use this framework to connect the droplet's statistics to the accompanying mean wave field.

% ----------------------------------------------
%         MEAN WAVE FIELD: FARADAY MODES
% ----------------------------------------------

\subsubsection{Contribution of Faraday modes}
\label{sec:Faraday_mean_wave}

The contribution of Faraday modes to the time-averaged wave field may be found by computing $$\langle \bm{\beta}\rangle = \lim_{N\rightarrow\infty}\frac{1}{N}\sum_{n=0}^{N-1}\bm{\beta}_n, $$
where $\vb_\mathrm{F}(t_n^+) = \bm{P}\bm{\beta}_n$ and we recall that $\eta$ is continuous across each impact.
We assume that the pilot-wave system (i) satisfies the ergodic hypothesis, which allows us to replace temporal averages with spatial averages, and (ii) exhibits a stationary probability distribution for the droplet position, denoted $p_s(\x)$\cite{DMB2018, DMW2020}. The apparent validity of these assumptions has been justified from various experimental and numerical observations\cite{DMB2018, DMW2020, DureyThesis, Saenz2018, NMB2017}. By applying the partial sum $N^{-1}\sum_{n = 0}^{N-1}$ to equation \eqref{eq:iterative_map_v2} and considering the limit as $N\rightarrow\infty$,  we deduce that
\begin{equation}
\label{eq:iterative_map_average_v1}
\langle\bm{\beta}\rangle = \bm{\Lambda}\langle\bm{\beta}\rangle - \bm{P}^{-1} \iint_\mathcal{D} p_s(\x) \bm{f}(\x) \,\mathrm{d}\x.
\end{equation}
By substituting in the definition of $\bm{f}(\x)$ (see equation \eqref{eq:f(x) forcing}) into \eqref{eq:iterative_map_average_v1} and rearranging, we conclude that
\begin{equation}
\label{eq:iterative_map_average_v2}
\langle\bm{\beta}\rangle = -\frac{mg T_\mathrm{F}}{\rho}(\bm{I} - \bm{\Lambda})^{-1}\bm{P}^{-1} \begin{pmatrix} \hat{\bm{p}}_s \\ \bm{0}\end{pmatrix},
\end{equation}
where $\hat{\bm{p}}_s$ is a column vector (of length $\mathscr{N}$) with elements 
$\iint_{\mathcal{D}}p_s(\x) \me^{-\mathrm{i}\bm{k}_j\cdot \bm{x}}\,\mathrm{d}\x$, corresponding to the discrete Fourier transform of the probability density function, $p_s(\x)$, over the domain, $\mathcal{D}$. Upon computing $\langle\bm{\beta}\rangle$, we may then determine the contribution of near-critical Faraday waves to the mean wave field from the product $\bm{P}\langle\bm{\beta}\rangle$.

% ----------------------------------------------
%         MEAN WAVE FIELD: LONG WAVES
% ----------------------------------------------

\subsubsection{Contribution of long waves}
\label{sec:long_mean_wave}

We proceed to develop an iterative map for the evolution of long waves, which we then use to determine the contribution of long waves to the mean wave field. During droplet flight, we deduce from equation \eqref{eq:QPE_system_forcing_long} that $\vb_\mathrm{L}(t)$ has linearly independent solutions
$$\vb_{\mathrm{L},j}(t) =  \me^{\mu_j t} \bm{q}_j\quad\mathrm{for}\quad j = 1,\ldots, 2\mathscr{N},$$
where the nonzero vector $\bm{q}_j$ and scalar $\mu_j$ satisfy the eigenvalue problem $\bm{A}\bm{q}_j = \mu_j \bm{q}_j$. We now determine an iterative map for $\vb_\mathrm{L}(t_n^+) = \bm{Q}\bm{\alpha}_n$ in terms of the matrix of eigenvectors $\bm{Q} = (\bm{q}_1, \bm{q}_2, \ldots)$ and the corresponding coefficients, $\bm{\alpha}_n$.
By following a framework similar to section \ref{sec:Faraday_map}, we deduce that the eigenmode coefficients, $\bm{\alpha}_n$, evolve according to the iterative map
$$\bm{\alpha}_{n+1} = \bm{D}\bm{\alpha}_n - \bm{Q}^{-1}\bm{f}_{n+1}, $$
where $\bm{D}$ is a diagonal matrix with diagonal elements $\me^{\mu_j T_\mathrm{F}}.$ This map is analogous to the map for the evolution of the Faraday modes (equation \eqref{eq:iterative_map_v2}). Finally, under the assumptions of ergodicity and the existence of a stationary probability density function for the droplet position, we follow a framework similar to section \ref{sec:Faraday_mean_wave} to determine the contribution of long waves to the mean wave field. Analogous to equation \eqref{eq:iterative_map_average_v2}, we deduce that the time-averaged eigenmode coefficients,
$$\langle\bm{\alpha}\rangle = \lim_{N\rightarrow\infty}\frac{1}{N}\sum_{n=0}^{N-1}\bm{\alpha}_n, $$
satisfy
\begin{equation}
\label{eq:iterative_map_average_long_wave}
\langle\bm{\alpha}\rangle = -\frac{mg T_\mathrm{F}}{\rho}(\bm{I} - \bm{D})^{-1}\bm{Q}^{-1} \begin{pmatrix} \hat{\bm{p}}_s \\ \bm{0}\end{pmatrix}.
\end{equation}

% ----------------------------------------------
%        MEAN WAVE FIELD: RECONSTRUCTION
% ----------------------------------------------

\subsubsection{Mean wave field structure}
\label{sec:reconstruct_mean_wave}

We determine the structure of the mean wave field using the approximate superposition of the contributions of Faraday modes and long waves, namely $\langle\eta\rangle \approx \langle\eta_\mathrm{F}\rangle + \langle\eta_\mathrm{L}\rangle$. We consider the contribution of different Faraday modes to $\langle\eta_\mathrm{F}\rangle$ (using $\langle\bm{\beta}\rangle$ defined in equation \eqref{eq:iterative_map_average_v2}),
where we order the modes so that $\gamma_\mathrm{F} = \gamma_1 \leq \gamma_2 \leq \gamma_3 \leq \ldots$ (and we neglect modes with critical threshold $\gamma_j < 0$, which are strongly damped). 
In particular, we approximate $\bm{p}_j(t)$ by neglecting higher harmonics in the wave oscillation (see equation \eqref{eq:p_def}) and consider $\x_j \sim \x_{j,0}$ and $\yb_j \sim \yb_{j,0}$; likewise, we approximate the decay rates using the asymptotic form $\lambda_j \sim \lambda_{j,1}(\gamma_j - \gamma)/g$. We also normalise each vector $\bm{p}_j(t_n)$ so that $\|\bm{p}_j(t_n)\| = 1$.
Two factors control the relative contribution of each Faraday mode to the mean wave field: (i) the proximity of the vibrational forcing to the mode's critical threshold; and (ii) the size of the projection of the probability density function, $p_s(\x)$, onto the Faraday mode. Owing to the complex structure of the Faraday wave field over the disordered topography, we find that $\langle \eta_\mathrm{F}\rangle$ may not be simply approximated by a small number of Faraday modes, with the mode weightings decreasing slowly with the mode number (see Fig.\,\ref{fig:sup_fig_faraday_modes}\fp{e}).
Finally, we compute $\langle\eta_\mathrm{L}\rangle$ (using $\langle\bm{\alpha}\rangle$ defined in equation \eqref{eq:iterative_map_average_long_wave}) with all $2\mathscr{N}$ eigenmodes included, with the results presented in Fig.\,\reply{\ref{fig:main_Fig_4}\fp{d}} of the main text.

% ----------------------------------------------

% ----------------------------------------------
%               QUANTUM
% ----------------------------------------------

\section{QUANTUM LOCALIZATION\label{sec:SI_quantum_localization}}

% Here is where we should first describe the nondimensionalization of Schrodinger's equation

% And then all the parameters of Anderson localizaiton:
% Weak-disorder regime
% Eigenmodes

\subsection{Schr\"odinger's equation }

We consider the time-dependent Schr\"odinger's equation in a 2D square domain, $\Omega$, of width $L$ with Dirichlet boundary conditions along the border, $\partial\Omega$. We  nondimensionalize the problem using $x_c=L$, $t_c=mL^2/\hbar$,  and $V_c=\hbar^2/mL^2$ as the characteristic scales for length, time and energy, respectively. Here, $m$ is the mass, and $\hbar$  the reduced Planck constant. In dimensionless form, the Schr\"odinger problem for the wave function, $\Psi(\bm{x}, t)$, thus becomes,
\begin{align}
\label{eq:Schrodinger}
\mathrm{i}  \frac{\partial\Psi}{\partial t} &=-\frac{1}{2} \nabla^2 \Psi+V(\bm{x}) \Psi,\\
\Psi &=0 \quad \text { on } \partial \Omega, \\
|\Psi(\bm{x}, 0)|^2&=C_0 \exp\left(-\frac{|\bm{x}|^2}{2\sigma^2}\right)\quad \text { on } \Omega,\label{eq:Schrodinger_ICs}
\end{align}
where $V(\bm{x})$ is the dimensionless potential, and $\Omega$ a square domain of width 1. The problem is initialized with a purely real Gaussian wave packet with standard deviation $\sigma$ that is normalized by $C_0$ such that the probably density function integrates to one, $\iint_{\Omega}|\Psi(\bm{x}, 0)|^2 \,\mathrm{d}S=1$.

Seeking separable solutions of the form $\Psi(\bm{x}, t)=\varphi(t)\psi(\bm{x})$, the general solution may written in the standard way as
\begin{equation}
    \Psi(\bm{x},t)=\sum_{n=1}^\infty c_n \me^{-\mathrm{i}E_n t}\psi_n(\bm{x}),
\end{equation}
where the energies, $E_n$, and normal modes, $\psi_n(\bm{x})$, are the eigenvalues and eigenfunctions of the time-independent Schr\"odinger equation,
\begin{equation}
    \label{eq:SchrodingerTimeIndep}
\mathcal{H}(\bm{x})\psi \equiv -\frac{1}{2} \nabla^2 \psi+V(\bm{x}) \psi=E\psi, 
\end{equation}
with the same boundary conditions. Here, $\mathcal{H}(\bm{x})$ represents the quantum Hamiltonian operator. The total energy may be computed   through the integral \(\ E=\iint_\Omega \Psi^\ast \mathcal{H}(\bm{x}) \Psi \,\mathrm{d}S\), where the asterisk denotes the complex conjugate. The evolution of the particle's mean-squared displacement, or second moment of the probability density function $|\Psi|^2$, is $\langle r^2(t)\rangle=\iint_\Omega|\Psi|^2r^2 \,\mathrm{d}S$, where $r = |\bm{x}|$ is the radial coordinate with respect to the origin of the coordinate system located at the center of the domain.

\subsection{Anderson localization}

Consider a classical particle evolving according to $m \ddot{\bm{x}}_\mathrm{p}(t)=-\nabla V(\bm{x}_\mathrm{p})$, where $V(\bm{x})$ is a random potential. The particle  is subject to weak disorder  when the total energy, $E=\frac{1}{2}m|\dot{\bm{x}}_\mathrm{p}|^2+V(\bm{x}_\mathrm{p})$, is larger than the characteristic potential energy of the background, $V_0$, i.e. $E/V_0>1$. The particle thus has sufficient energy to explore the entire potential,  the process during which the particle's trajectory becomes deflected in random directions,  leading to diffusive motion in the long-time limit in two (or higher) dimensions\cite{bechinger2016active}.

Anderson localization describes the fundamentally different behavior of quantum particles in the equivalent situation. Owing to their wave-particle duality, quantum particles exhibit a remarkable phenomenon wherein even a weak disordered potential causes them to spontaneously halt, in stark contrast to classical expectations\cite{MullerDelande2011Andersontheory}. This absence of diffusion is observed  for a sufficient degree of randomness; the 
emergent eigenmodes become exponentially localized (Fig.\,\ref{fig:sup_fig_eigenfunctions_schrodinger}) even when the total energy of the particle is larger than the potential energy of the heterogeneous background\cite{filoche2012universal}. 
In the quantum regime, the  conditions for weak disorder \cite{MullerDelande2011Andersontheory} are $E_\zeta E/V_0^2 > 1$ and $E_\zeta/V_0> 1$, where $E_\zeta=1/\zeta^2$ is the dimensionless energy associated to the potential's correlation length, $\zeta$.  
The correlation length, $\zeta$, and characteristic magnitude of the random potential, $V_0$, are computed through the average correlation function  $\langle V(\bm{x}^{\prime}) V(\bm{x}^{\prime}+\bm{x})\rangle=V_0^2 f(r)$, where $f(r)$ is a function with $\max f=1$ at the origin\cite{MullerDelande2011Andersontheory}. For our potential composed of squares tiles of constant width and random height, $f(r)$ decays linearly near the origin. We thus approximate $f(r)\approx -V_0^2(r/\zeta-1)$ in the interval $0\leqslant r\leqslant W$ to compute the correlation length, $\zeta$. As expected for our random potential, we find that the correlation length roughly corresponds to the tile width, $\zeta\approx W$, in all realizations. 

Observation of Anderson localization is also dependent on the system dimension. The scaling theory of localization specifies that all states become localized in 2D while an energy threshold, or `mobility edge', separates localized and non-localized states in higher dimensions\cite{abrahams1979scaling}. Moreover, the quantum system must be tuned into a scattering regimen for which the product $kl$ should be of the order of unity in 2D (and smaller in higher dimensions), where $k$ is the de Broglie wavenumber and $l$ the mean free path\cite{MullerDelande2011Andersontheory}. Assuming that the mean free path is roughly comparable to the tile width, $l\approx W$, we obtained $kl=0.9 $ for the simulations presented in Fig.\,\ref{fig:main_Fig_2}.

%Observation of Anderson localization also requires the system to be in the scattering regime. In two dimenstions,

%\todo{[Frane: The localization length is a parameter that indicates [CITE dominique?] how large should be the size of the domain considered such that localization can occur. It also represents quantitatively the characteristic length  over which the steady state localized wave packet should extend. The scaling theory of localization \cite{abrahams1979scaling} specifies that for noninteracting particles in 2D all states become localized and that the localization length is given by $\xi_{\mathrm{loc}}
%\left.\sim L_{0} \exp \left\{2 g_{0} / \pi\right)\right\}=l \exp \left\{\pi k l / 2\right\}
 %\gg l $]. The general simplified criterion is that to observe localization the product $kl$ should be chosen as close to the unity as possible. Where $l$ is the mean free path, $k=2\pi/\lambda$ and $\lambda$ is De Broglie wavelength associated to the particle [should this be explained in more detail ? meaning the relationship of lambda to $V_0$ and that $l=1/\textrm{\#tiles}$]}

%\todo{We also need to add the localization length stuff?}

% \textcolor{cyan}{Reminders regarding Fig 2:\\ Need to say that Anderson localization is surprising due to the emergence of bound eigenmodes with exponential decay rather than scattering states, which you would expect when the particle energy is large in comparison to the characteristic potential.}

\subsubsection{Localized eigenfunctions\label{sec:localized_eigenfunctions}}

For walking droplets above variable bottom topography, deeper and shallower areas act as regions of lower and higher potential, respectively\cite{Saenz2019,Cristea-platon2018}. To investigate qualitatively the quantum eigenmodes for a random medium of the same form as the walker's bottom topography in the experiment, we thus define the dimensionless potential for the Schr\"ondinger problem \eqref{eq:SchrodingerTimeIndep} as
\begin{equation}
V(\bm{x})=-K \left( H(\bm{x})-\max H(\bm{x}) \right),
\label{eq:dimless_potential}
\end{equation}
 where $K=\kappa/V_c$ and $H(\bm{x})$ have dimensions of 1/length and length, respectively. Here, $H(\bm{x})$ is the random liquid depth distribution used in the experiments, and $\kappa$ the dimensional scaling constant presented in the main text, which has dimensions of energy/length. We note that \eqref{eq:dimless_potential} is equivalent to 
\begin{equation}
V(\bm{x})=K \Delta \hat{H}(\bm{x}),
\label{eq:dimless_potential2}
\end{equation}
where $\Delta \hat{H}(\bm{x})$ are the random heights drawn from the uniform distribution in the interval $[0, 0.6]$\,mm.

\begin{table}[]
\begin{tabular}{|c|c|c|c|c|c|}
\hline
$K$\,(mm\textsuperscript{-1})     & $V_0$  & $E_1$ & $E_\zeta$  & $E_\zeta E_1/V_0^2$ & $E_\zeta /V_0$ \\ \hline
27    & 6.16   & 18.3  & 135.6  & 65.3   &  22.0     \\
170   & 38.8   & 59.7  & 135.6  & 5.38   &  3.50     \\
237   & 54.1   & 76.5  & 135.6  & 3.55   &  2.51     \\
350   & 79.8   & 98.5  & 135.6  & 2.09   &  1.70     \\
1300  & 296.6  & 192.0 & 135.6  & 0.30   &  0.46     \\ \hline
\end{tabular}
\caption{Dimenionless energies and energy ratios for the first eigenfunction (ground state, $E_1$) of the time-independent Schr\"ondinger equation \eqref{eq:SchrodingerTimeIndep} shown in Fig.\,\ref{fig:sup_fig_eigenfunctions_schrodinger}\fp{a} for different values of the scaling factor $K$ in the random potential \eqref{eq:dimless_potential}.
\label{tab:energies}}
\end{table}

\begin{figure*}
\includegraphics[scale=1]{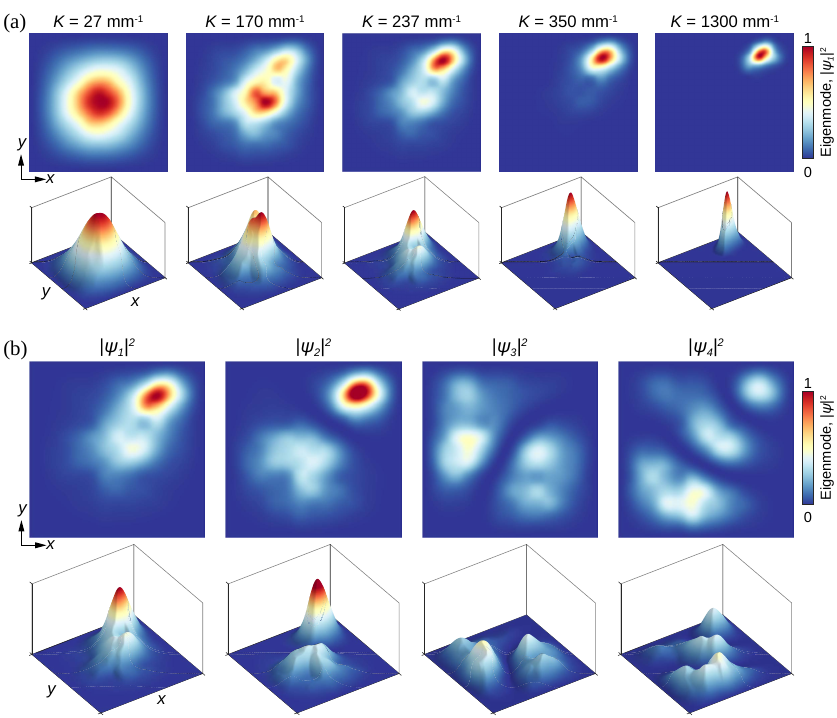}
\caption{\textbf{Eigenfunctions of Schr\"odinger's equation.} 
(a) Ground states , $|\Psi_1|^2$, arising from  Schr\"odinger's equation \eqref{eq:SchrodingerTimeIndep} with a potential of the same form as the walker's bottom topography \eqref{eq:dimless_potential} for different values of re-scaling constant $K$. The associated energies are listed in Table \ref{tab:energies}. 
(b) First four eigenfunctions for the potential with $K=237$\,mm\textsuperscript{-1}, whose dimensionless energies are $E_1=76.5$,  $E_2=86.9$, $E_3=102.2$ and   $E_4=107.4$, respectively. The first eigenfunction, $\Psi_1$, is the mode presented in Fig.\ref{fig:main_Fig_1}.
%\todo{Explain nomralization}
}
\label{fig:sup_fig_eigenfunctions_schrodinger}
\end{figure*}

We discretize the time-independent  Schr\"ondinger equation \eqref{eq:SchrodingerTimeIndep} using a finite-difference method with a central difference for the space derivatives, and Dirichlet conditions along the border. In matrix form, the problem thus becomes $\bm{M}\bm{\psi}_n=E_n\bm{\psi}_n$ (for $n=1,2,...$), where the vector $\bm{\psi}_n$ is the discretized eigenfunction with energy $E_n$, and $\bm{M}$ is a matrix of coefficients. We discretize the domain, which is composed of $13\times 13$ tiles, using $256\times256$ points ($\sim 20$ points per tile), and verified that using $512\times512$ points ($\sim 39$ points per tile) yielded  virtually the same results. The emergent eigenfunctions for several values of the re-scaling constant $K$ are presented in  Fig.\,\ref{fig:sup_fig_eigenfunctions_schrodinger}\fp{a}, and the associated energies are listed in Table \ref{tab:energies}. The mode presented in the main text (Fig.\,\ref{fig:main_Fig_1}) corresponds to $K=237$\,mm\textsuperscript{-1}. Higher-energy modes for the same value of $K$ are presented in Fig.\,\ref{fig:sup_fig_eigenfunctions_schrodinger}\fp{b}.

%\markblue{Frane reported: Correlation length $\zeta=0.0888$, characteristic potential energy $V_0=0.6946$, energy of first mode $E=0.97$, correlation energy should be $E_\zeta=126.81$. The weak disorder limits are thus $E_\zeta E/V_0^2=255$ and $E_\zeta/V_0=182$.}

\subsubsection{Absence of diffusion}

To demonstrate the absence of diffusion of quantum particles in disordered media, we performed simulations of the time-dependent Schr\"odinger's equation \eqref{eq:Schrodinger} in a domain larger than that described in Fig.\,\ref{fig:main_Fig_1}. Specifically, we  increased the number of tiles to $37\times 37$, and readjusted the re-scaling constant $K=3200$\,mm\textsuperscript{-1} in \eqref{eq:dimless_potential2} to obtain energy ratios similar to those corresponding to the eigenmode presented in Fig.\,\ref{fig:main_Fig_1}. We replaced the regularly spaced pillars, which do not play a significant role, with tiles of a random height drawn from the same uniform distribution. We averaged over 20  random realizations of the heterogeneous media. To facilitate the averaging, for each random realization, we shifted the potential in the $xy$-plane so that the peak of the ground state was located at the center of the domain. For the initial Gaussian wave packet \eqref{eq:Schrodinger_ICs}, we selected a dispersion of  $\sigma= 1/\left(12\sqrt{2}\right)$ and the normalization constant was $C_0=2.114 \times 10^{-3}$. 
The resulting average energy ratios were $E_\zeta E/V_0^2=3.47\pm0.30$ and $E_\zeta/V_0=2.18\pm0.07$, where the interval represents the standard deviation and $E$ is the total energy.
We integrated Schr\"odinger's equation \eqref{eq:Schrodinger}  numerically using the Crank-Nicholson scheme to advance time. For the simulations presented in the main text, we used a spatial resolution of $N\times N=222\times222$ points ($\sim6$ points per tile), and a temporal resolution of $\Delta t=1/(2 N^2)$. We performed numerical tests with other resolutions (4, 5, 7, and 8 points per tile) to ensure that our results were independent of the spatio-temporal discretization. As the resolution was increased, the final MSD converged to the same value. For the resolution selected for the final simulations, the MSD differed by only $0.32\%$ with respect to that of the highest resolution.  
The temporal evolution of average wave packet and MSD are presented in Fig.\,\ref{fig:main_Fig_2}. The wave function effectively freezes, i.e.~the MSD saturates, well before reaching the maximum allowed by the size of the domain, specifically $(L/2)^2=0.25$.

%\todo{Frane:[localization length] do we need statistics for the localization length and $kl$ product? Pedro: I'm not sure.}

%\subsection{Numerical simulations (Integrate in previous sections and remove)}

%\todo{The numerical scheme considered is Crank-Nicholson first order in time and space. It is important to note that this method is computationally expensive, but it is more precise and more stable than other low-order time-stepping methods. It calculates the time derivative with a central finite differences approximation. The simulation was stopped once the initial condition would interact with the boundary so boundary conditions would minimally affect the simulation \markblue{(We need to know how we are going to evolve in time asap.)}. The energy of the eigenmodes is numerically computed solving equation \eqref{eq:SchrodingerTimeIndep}. The energy of the wavefunction if computed via \(\ E=\int_\Omega \Psi^\ast \mathcal{H}(\bm{x}) \Psi \,\mathrm{d}S\), where the asterisk denotes the complex conjugate. The Hamiltonian is discretized numerically to first order for this calculation.}

% ----------------------------------------------
%               SIMULATIONS
% ----------------------------------------------

\section{SIMULATIONS
\label{sec:SI_simulations}}

\subsection{Walking droplets\label{sec:SI_walker_simulation}}

% ----------------------------------------------
%              WALKING DROPLETS
% ----------------------------------------------

We perform numerical simulations of walking droplets using the pilot-wave model of Faria\cite{Faria2016}, which couples a variable-topography waves model \eqref{eq:QPE} with the trajectory equation\cite{MB2013b} for the droplet \eqref{eq:drop_trajectory}, as described in section \ref{sec:pilot-wave_model}.
Between droplet impacts, which are modelled as delta functions in time and space, the droplet evolution \eqref{eq:drop_trajectory} is solved analytically, while the waves model \eqref{eq:QPE} is solved numerically in a doubly-periodic square domain using a pseudospectral method in space, and  a fourth-order Runge-Kutta scheme for the time integration\cite{Faria2016}. Each droplet impact is resolved analytically by applying appropriate jump conditions\cite{Faria2016, DM2017} in the droplet's velocity \eqref{eq:drop_trajectory_delta_jump}, and fluid velocity potential \eqref{eq:f(x) forcing}.% \textcolor{red}{Matt: we need to state the value of the impact phase for each simulation.}

\subsubsection{Faraday threshold\label{sec:faraday_threshold}}

\begin{figure}[]
\includegraphics[scale=1]{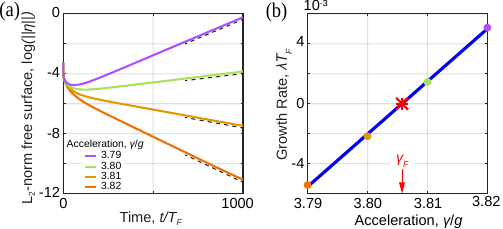}
\caption{\textbf{Numerical estimation of the Faraday threshold.} 
(a) Temporal evolution of  the $L_2$-norm of the free surface, $\| \eta \|$, for different vibrational forcings, $\gamma$.
(b) Calculation of the Faraday threshold, $\gamma_\mathrm{F}$, by applying a linear least-squares fit to the time exponential growth/decay rates, $\lambda$.
} 
\label{fig:sup_fig_faraday_threshold}
\end{figure}

The Faraday threshold, $\gamma_\mathrm{F}$, is the critical vibrational acceleration, $\gamma$, above which perturbations in the bath's fluid rest state grow. The growth is exponential in time for $\gamma > \gamma_\mathrm{F}$, with exponential decay for $\gamma < \gamma_\mathrm{F}$ (see section \ref{sec:Floquet_theory} and Fig.\,\ref{fig:sup_fig_faraday_threshold}\fp{a}).
%amplitude $\eta_{\text{max}}$ decays exponentially \todo{(check)} in time
%Since the fluid evolution equations considered in \eqref{eq:QPE} form a linear wave system \textcolor{red}{(why is this relevant?)}, the amplitude $\eta_{\text{max}}$ of the wave surface exhibits linear \textcolor{red}{(really? I would have thought exponential)} growth in time for accelerations $\gamma$ satisfying $\gamma>\gamma_F.$ 
Near to the Faraday threshold, we expect the wave growth/decay rate, $\lambda(\gamma)$, to satisfy $\lambda(\gamma) \sim \lambda_0(\gamma - \gamma_\mathrm{F})$, with $\lambda_0 > 0$ so that $\lambda > 0 $ for $\gamma > \gamma_\mathrm{F}$ and $\lambda < 0$ for $\gamma < \gamma_\mathrm{F}$ (see section \ref {sec:decay_rates}).
Consequently, we may obtain a numerical estimation of the Faraday threshold for each realization of the random topography by exciting a disturbance in the wave surface and measuring the wave growth/decay for different acceleration values, $\gamma$.

Specifically, we characterise the evolution of the free surface in terms of the $L_2$-norm of the free surface, $\| \eta \|$, computed over the spatial domain. For a given value of $\gamma$, we initialise each simulation by 
%We compute the Faraday threshold by 
exciting a zeroth-order Bessel function in the deepest region of the topography and recording $\|\eta\|$ over 1000 Faraday periods. We then numerically estimate the long-time exponential growth/decay rate, $\lambda(\gamma)$ (Fig.\,\ref{fig:sup_fig_faraday_threshold}\fp{a}). Upon repeating this process for several values of $\gamma$ close to the instability threshold, we estimate the parameters $\lambda_0$ and $c = -\lambda_0\gamma_\mathrm{F}$ with a linear least-squares fit, $\lambda \sim c + \lambda_0\gamma$ (Fig.\,\ref{fig:sup_fig_faraday_threshold}\fp{b}). Upon computing $c$ and $\lambda_0$, the Faraday threshold is then $\gamma_\mathrm{F} = -c/\lambda_0$.  
%The critical value $\gamma_F$ is then defined to be the largest value of $\gamma$ such that $\eta_{\text{max}}<10^{-6}.$ \textcolor{red}{Matt: I suggested a more precise method (but it wasn't adopted) -- if we re-run any simulations then I think we should use the new method instead. The current method may not be not sufficiently precise when operating close to the Faraday threshold.}

\subsubsection{Simulation of the experiments\label{sec:walker_simulation_experiment}}

\begin{figure*}
\includegraphics[scale=0.9]{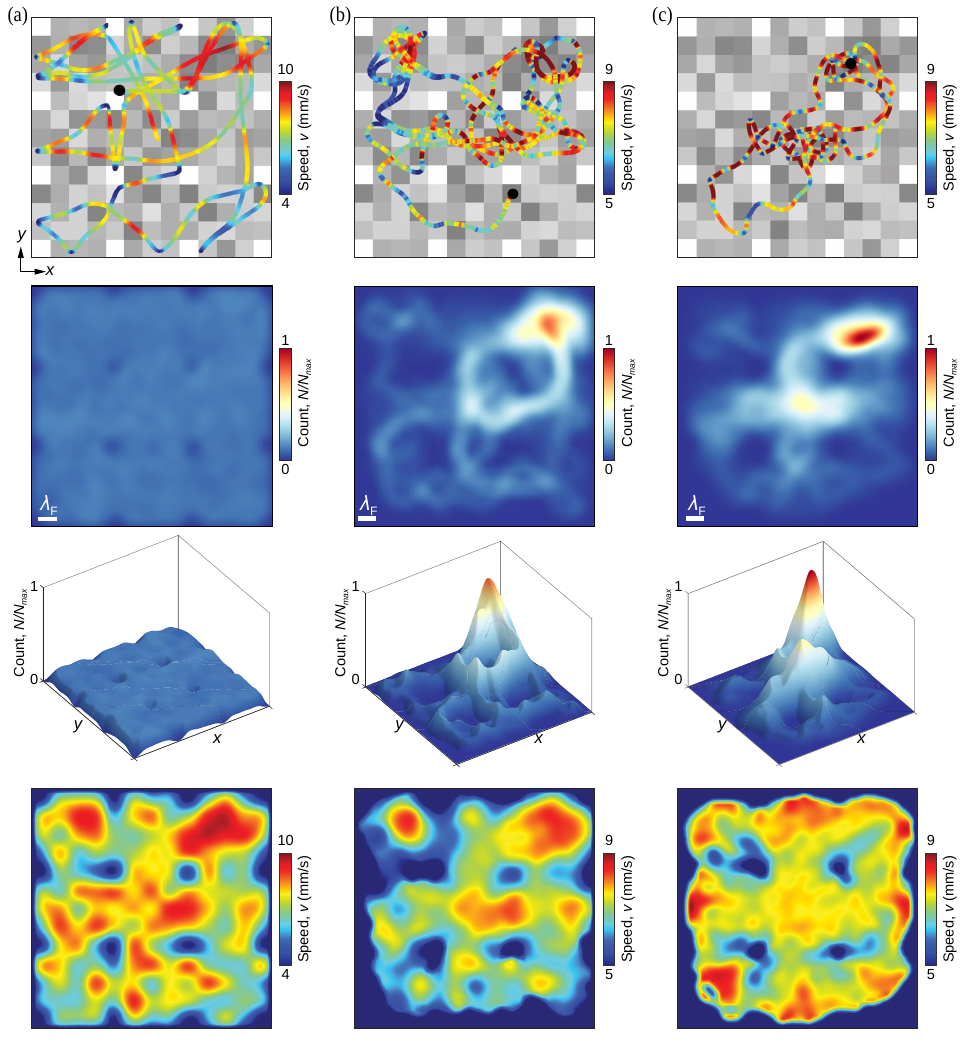}
\caption{\textbf{Emergent localized statistics in the walker system.} Comparison of the speed-colored trajectory (first row), position histogram (second and third rows), average speed map (fourth row) for (a) a waveless particle (see section \ref{sec:SI_waveless_simulations}), and a walker according to (b) experiments (see section \ref{sec:SI_experiment}), and (c) simulations (see section \ref{sec:SI_walker_simulation}). The position histograms, $N(\bm{x})$ (computed using 3 bins per wavelength in each direction), were first normalized such that
$\iint_\Omega N(\bm{x}) \,\mathrm{d}S=1$, and later re-scaled by the maximum of the three histograms, $N_\mathrm{max}$. The average speed maps were computed using a $78\times78$ regular grid and a Gaussian-weighted average, with standard deviation $\sigma=\lambda_\mathrm{F}/4$, applied to data points within $\lambda_\mathrm{F}/4$ distance away from each grid point.
\label{fig:sup_fig_waveless_experiment_simulations}}
\end{figure*}

To test whether the pilot-wave model \eqref{eq:QPE}-\eqref{eq:drop_trajectory} captures the walker localization observed in experiments, we integrated the governing equations numerically using a square domain of size $L\times L =19\lambda_\mathrm{F}\times19\lambda_\mathrm{F}$, in which the random topography of size $13\lambda_\mathrm{F}\times13\lambda_\mathrm{F}$ was surrounded by a damper (i.e.\ a shallow fluid layer) of width $3\lambda_\mathrm{F}$.  Preliminary tests showed that the influence of the random topography on the droplet dynamics in the simulations was weaker than that in the experiments. Maintaining the same base depth as in the experiments, we thus re-scaled the random fluid depths in the simulations as $H(\bm{x})=H_0 - (5/3)\Delta H(\bm{x})$, including the height of the damper (Fig.\,\ref{fig:sup_fig_border}\fp{a}). The problem was discretized using a spatial resolution of $N\times N=204\times204$ points ($\sim11$ points per wavelength), and a temporal resolution of $\Delta t=T_\mathrm{F} L/(10 N \lambda_\mathrm{F})$, where  $T_\mathrm{F}=2/f$ is the Faraday period. Numerical tests were performed to ensure that the spatio-temporal resolution was adequate to render discretization-independent results (see section \ref{sec:simulaitons_numerical_convergence}). 
We also performed tests with different widths of the wave damper to ensure that the periodic boundary conditions had no influence of the emergent statistics
(see section \ref{sec:sec:walker_damper}). 

The fluid properties and droplet radius were the same as in the experiments. The effective kinematic viscosity, $\nu_e=0.828\nu$, was chosen to  match the experimental stability threshold, $\gamma_\mathrm{F}$\cite{
MGNB2015,Faria2016}. The simulations were performed at $\gamma/\gamma_\mathrm{F}=99.89\%$, where $\gamma_\mathrm{F}=3.806g$ is the numerical threshold. We observed that the strength of the droplet localization was significantly influenced by the coefficient of restitution, which we set to  $c_4=0.05$. We note that this value is lower than the typical values considered in the literature, $0.13\leqslant c_4\leqslant 0.33$  \cite{MGNB2015, MB2013b}.
We tuned the impact phase, $\theta_I=1.418\pi$\,rad, to match the average speed in the simulations to that in the experiments.  We simulated 36 walkers initialized with random conditions, each for 21,000 impacts, or Faraday periods, $T_\mathrm{F}$. The simulations thus yielded a total of 756,000 impacts, matching the duration of our experiments. The droplet position was saved at each impact. We also confirmed the ergodicity of the system (see section \ref{sec:ergodicity_check}).

\begin{figure}[]
\includegraphics[scale=1]{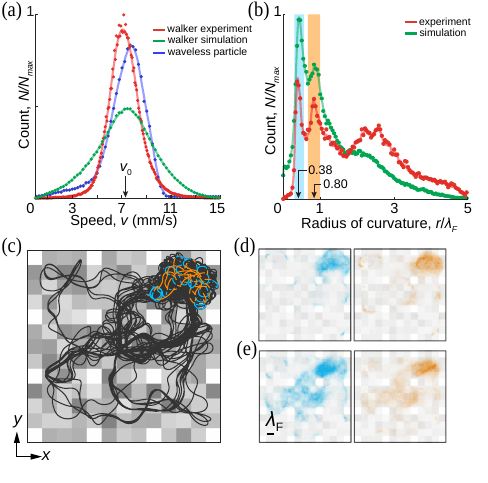}
\caption{\textbf{Walker dynamics in experiments and simulations.} Histograms of the instantaneous (a) droplet speed, and (b) radius of curvature of the droplet trajectory. Each histogram, $N(s)$, was normalized such that
$\int_{-\infty}^{+\infty} N(s) \,\mathrm{d}s=1$, and later re-scaled by the maximum of the histograms, $N_\mathrm{max}$. Preferred radii of curvature around $r/\lambda_\mathrm{F}\sim 0.38$ and 0.80 arise in both experiments and simulations. (c) To examine if there is a spatial correlation between the preferred radii and the localized statistics, we segment the droplet trajectory according to the local radius of curvature into two intervals, $0.30\leqslant r/\lambda_\mathrm{F}\leqslant0.56$ (blue area in b) and $0.67\leqslant r/\lambda_\mathrm{F}\leqslant 1$ (orange area in b), containing the preferred radii. By computing the spatial histograms of the segmented trajectories for the (d) experiments, and (e) simulations,
we observe that the loops with preferred radii mainly occur in the localization region.
}
\label{fig:sup_curvature_maps}
\end{figure}

The results of our simulations are presented in Fig.\,\ref{fig:sup_fig_waveless_experiment_simulations}\fp{c}. The position histogram features localized statistics that agree well with the experiments. We note that the average speed map depicts some differences with respect to the experimental map, which indicates that the walker localization is not significantly influenced by the details of the instantaneous speed. Indeed, comparing the speed histograms  revealed a broader distribution in the simulations (Fig.\,\ref{fig:sup_curvature_maps}\fp{a}). The differences may be rooted in the simplifying assumptions adopted for the derivation of theoretical pilot-wave model \eqref{eq:QPE}-\eqref{eq:drop_trajectory}, including the modelling of the droplet bounces as point impacts and the prescription of purely sub-harmonic vertical dynamics. For example, we observed in experiments that \reply{occasionally} the walker experienced chaotic changes in its vertical phase near the localization region. \reply{More subtle phase shifts may also potentially be induced  by random wave reflections from topographical features, which can alter the wave amplitude at the droplet's position and thus its bouncing phase. While these effects, which are not accounted for in our simulations, may contribute to some differences in the speed distributions,} the fact that the simplified model captures the walker localization suggests that these are unessential details.

We also compared the curvature of the walker trajectories in experiments and simulations. Notably, the preferred radii observed in the experiments were also apparent in the simulations (Fig.\,\ref{fig:sup_curvature_maps}\fp{b}). To demonstrate that the preferred radii occur near the localization region, where the droplet executes many loops, we segmented the droplet trajectories according to the local radius of curvature into two intervals, $0.30\leqslant r/\lambda_\mathrm{F}\leqslant0.56$ and $0.67\leqslant r/\lambda_\mathrm{F}\leqslant 1$, containing the preferred radii (Fig.\,\ref{fig:sup_curvature_maps}\fp{c}). By computing the position histogram for each segment, we observed that the loops with preferred radii indeed occur predominately near the localization region  (Fig.\,\ref{fig:sup_curvature_maps}\fp{d-e})

\begin{figure}[]
\includegraphics[scale=0.9]{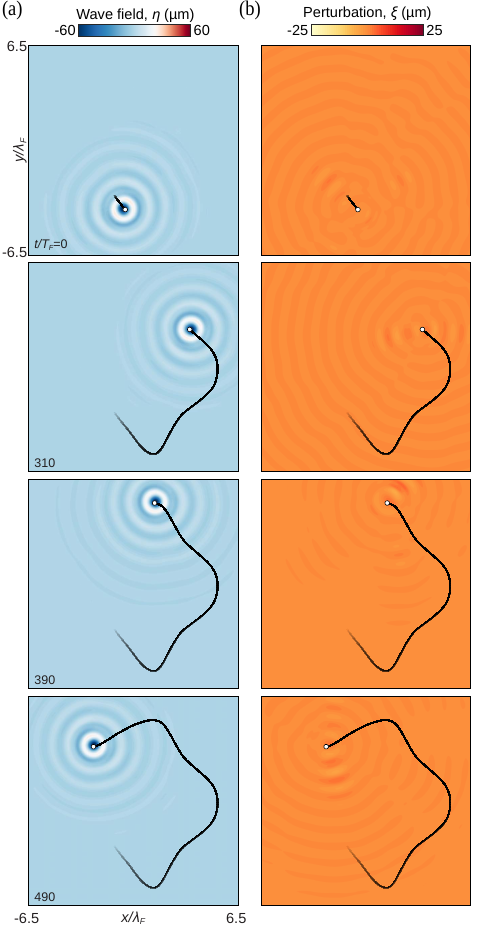}
\caption{\textbf{Walker wave field at low memory.} Repeating the simulations described in Fig.\,\ref{fig:main_Fig_3} at low memory, $\gamma/\gamma_\mathrm{F}=83.0\%$, reveals (a) a less extensive and more circularly-symmetric wave field, and (b) significantly weaker topography-induced wave perturbations. Thus, the influence of the submerged topography on the walker dynamics diminishes as the memory is reduced. In this simulations, the average walker speed was maintained the same as in Fig.\,\ref{fig:main_Fig_3} by readjusting the impact phase, $\theta_I=1.367\pi$\,rad.
\label{fig:sup_fig_LMwave}}
\end{figure}

\subsubsection{Wave field\label{sec:wave_field}}

To rationalize the mechanism responsible for the walker localization, we present a series of wave-mediated interactions between the droplet and submerged topography in Fig.\,\ref{fig:main_Fig_3}. Experimentally, the wave field was captured using the semi-reflective mirror set-up described in section \ref{sec:experimental_set-up}. The video recording was strobed at the Faraday frequency (35 frames per second) and the acquisition phase (relative to the external driving) was tuned to record the waves near their maximum amplitude. The simulated wave field given by the quasi-potential model \eqref{eq:QPE} was similarly saved with the Faraday frequency near its maximum amplitude, which roughly corresponded to a phase 26\% of the Faraday period after each impact.

To illustrate further the role of the memory, we compare the wave field at high memory, $\gamma/\gamma_\mathrm{F}=99\%$ (Fig.\,\ref{fig:main_Fig_3}), and low memory $\gamma/\gamma_\mathrm{F}=83\%$ (Fig.\,\ref{fig:sup_fig_LMwave}), using simulations, which allow us to maintain the same walker speed by readjusting the impact phase. At low memory, the waves decay significantly faster in time. The resulting walker wave field is thus less extensive and more weakly influenced by the topography (Fig.\,\ref{fig:sup_fig_LMwave}). As the memory decreases, the force on the walker thus becomes more local (i.e.~less sensitive to the prior path or spatially distant features), approaching the behavior of a `waveless' particle (see section \ref{sec:SI_waveless_simulations}) whose evolution is purely local (i.e.\ the force acting on the particle depends entirely on its current position).

\begin{figure}[]
\includegraphics[scale=1]{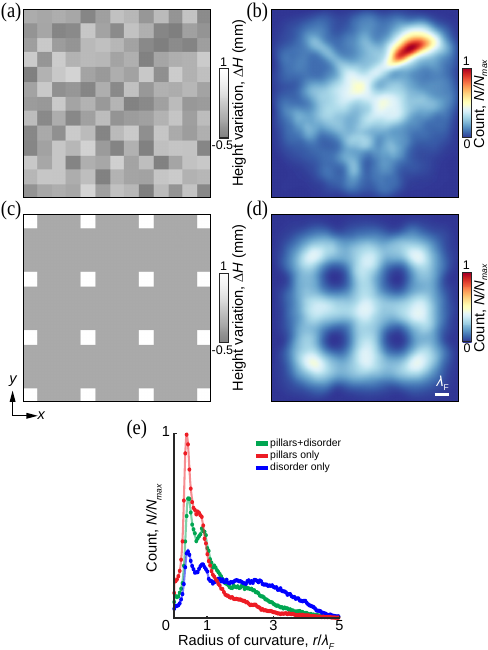}
\caption{\textbf{Walker statistics in domains without regularly-spaced pillars or disorder.} Walker simulations with (a) a fully disordered submerged topography  result in localized statistics as shown by (b) the \reply{droplet's} position histogram. If the topography is ordered and only includes regularly-spaced pillars (c), the position histogram remains symmetric (d). (e) Comparison of the curvature distributions. 
\label{fig:sup_fig_pillars}}
\end{figure}

\subsubsection{Influence of pillars\label{sec:pillars}}

To demonstrate that the inclusion of regularly spaced pillars  (tiles with the same height above the base level in Fig.\,\ref{fig:main_Fig_1}\fp{c}) plays no significant role in the localization problem, we repeated the simulations described in section \ref{sec:walker_simulation_experiment} without them. Specifically, we made the height of each pillar random according to the same uniform distribution used for the rest of the tiles. The height of the remaining tiles was maintained as before. The resulting submerged topography is presented in Fig.\,\ref{fig:sup_fig_pillars}\fp{a}. The vibrational acceleration was readjusted to maintain the same memory, $\gamma/\gamma_\mathrm{F}$, with the new Faraday threshold being $\gamma_\mathrm{F}=3.783g$. We readjusted the impact phase, $\theta_I=1.419\pi$\,rad, to recover the same average speed. The emergent position histogram is presented in Fig.\,\ref{fig:sup_fig_pillars}\fp{b}, which features localized statistics in the same region as in the case with pillars shown in Fig.\,\ref{fig:sup_fig_waveless_experiment_simulations}\fp{c}.

To further demonstrate that the localized statistics are caused by the random topography, we also repeated the simulations with a topography that only included the regularly-spaced pillars. We thus set the depth of the fluid for the rest of the tiles to be the same as the base depth $H_0$ (Fig.\,\ref{fig:sup_fig_pillars}\fp{c}). The new Faraday threshold was $\gamma_\mathrm{F}=3.825g$, and the impact phase was reset to its original value $\theta_I=1.418\pi$\,rad. Without randomness, the emergent histogram shows no localization (Fig.\,\ref{fig:sup_fig_pillars}\fp{d}). The emergent curvature distributions are presented in Fig.\,\ref{fig:sup_fig_pillars}\fp{e}, which illustrate that the experimental topography with disorder between the pillars is an intermediate configuration between the regular topography with pillars and the fully disordered topography.

\reply{\subsubsection{Role of disorder\label{sec:disorder}}}

\begin{figure*}
\includegraphics[scale=1]{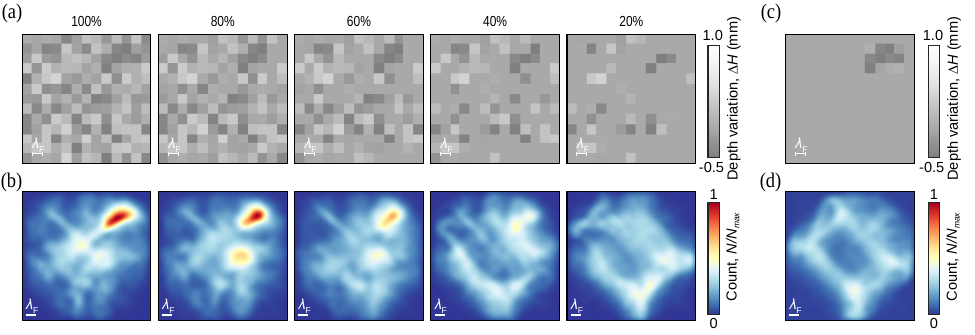}
\caption{\reply{\textbf{Influence of disorder.} Walker simulations with (a) a decreasing percentage of randomly selected tiles with heterogeneous heights along with (b) the resulting position histograms. (c) A specific configuration wherein disorder is retained solely in the region where localization emerges with a completely random topography, along with (d) its position histogram, which does not exhibit localization.} 
\label{fig:sup_fig_disorder}}
\end{figure*}

\reply{To examine how disorder influences the formation of localized statistics in walking droplets, we conducted a series of simulations similar to those described in  \ref{sec:walker_simulation_experiment}, varying the degree of disorder in the submerged topography. 
Having shown that the pillars are not relevant for localization in section \ref{sec:pillars}, we took the configuration depicted in Fig.\,\ref{fig:sup_fig_pillars}\fp{a} as the basis for this exploration. We then progressively reduced the level of disorder by randomly selecting a percentage of tiles and setting their depth equal to the base depth (Fig.\,\ref{fig:sup_fig_disorder}\fp{a}). A secondary localization peak appeared while the degree of disorder was relatively high; however, any indications of localization vanished when roughly half of the topography became ordered (Fig.\,\ref{fig:sup_fig_disorder}\fp{b}). 
Furthermore, we performed an additional set of simulations wherein disorder was preserved exclusively in the region where localization had been observed with a completely random topography, setting the depth elsewhere in the bath equal to the base depth (Fig.\,\ref{fig:sup_fig_disorder}\fp{c}). Despite the disordered region now being the deepest area in the domain, the localization disappears from the position histogram (Fig.\,\ref{fig:sup_fig_disorder}\fp{d}). Comprehensive tests were performed to ensure that these results were independent of the initial conditions. These results thus indicate a nontrivial relationship between disorder and long-range effects in the emergence of localization with walking droplets.
}

\

\subsubsection{Wave damper\label{sec:sec:walker_damper}}

\begin{figure}[]
\includegraphics[scale=1]{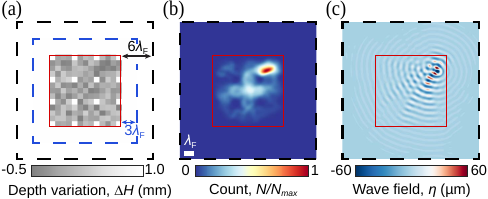}
\caption{\textbf{Influence of the wave damper in the walker simulations.} 
(a) Re-scaled height about the base depth, $H_0=1.85$\,mm, of the submerged topography used in the simulations described in section \ref{sec:walker_simulation_experiment}. The heterogeneous region is surrounded by a wave damper of width $3\lambda_\mathrm{F}$, which was not shown in previous figures. Dashed lines denote the domain boundary along which periodic boundary conditions are prescribed. 
Simulations with a wider border, $6\lambda_\mathrm{F}$, result in the same position histogram (b). (c) Visualization of the walker wave field demonstrates that the waves are indeed adequately damped above the border before they may reappear in the domain of interest due to the periodic boundary conditions. The wave field is shown at its maximum amplitude (section \ref{sec:wave_field}).
\label{fig:sup_fig_border}}
\end{figure}

We performed a series of simulations to investigate the influence of the wave damper surrounding the heterogeneous region. The submerged topography used for the reference simulation described in section \ref{sec:walker_simulation_experiment} is presented in Fig.\,\ref{fig:sup_fig_border}\fp{a}, for which the damper has a width of $3\lambda_\mathrm{F}$ (the effective width of the damper is $6\lambda_\mathrm{F}$ due to the periodic boundary conditions). To ensure that the waves were adequately damped above the border, we repeated the simulation with a damper twice as wide, as shown in Fig.\,\ref{fig:sup_fig_border}\fp{a}. We recomputed the Faraday threshold, $\gamma_\mathrm{F}$, to ensure that the prescribed memory, $\gamma/\gamma_\mathrm{F}$, was the same as in the reference case. Our simulations with the wider damper yielded virtually the same position histogram (Fig.\,\ref{fig:sup_fig_border}\fp{b}). Moreover, we visualized the walker wave field to check that the waves were indeed damped above the border, thereby not `wrapping around' the periodic domain (Fig.\,\ref{fig:sup_fig_border}\fp{c}). Our tests thus demonstrate that the wave damper is appropriate to ensure that the periodic boundary conditions in our simulations do not influence the walker dynamics.

\subsubsection{Absence of diffusion\label{sec:absence_of_diffusion_walker}}

To demonstrate how the localized position statistics may lead to an absence of diffusion in the walker system, we performed simulations of walking droplets in domains larger than those accessible experimentally.  Specifically, we considered \reply{8} different random realizations of the potential in a domain with \reply{$36\times36$} tiles of the same width as in the experiments. Each realization was explored by \reply{400} walkers for \reply{$2\times10^4$} bounces. The base depth was tuned to $H_0=1.45$\,mm. The regularly-spaced pillars from the experiment were excluded; the heights about the base depth of all the tiles were drawn from \reply{a re-scaled distribution $H(\bm{x})=H_0 - (4/3)\Delta H(\bm{x})$, chosen to show the full transition from diffusion to localization when sweeping in memory.} 
We also removed the wave damper. Guided by preliminary simulations, we shifted the random patterns on the \textit{xy}-plane to center the localization region on the computation domain, and so facilitate the spatial averaging of the position histograms. 
The problem was solved numerically using \reply{$\sim7$} points per wavelength and temporal resolution as in the simulations of the experiments (see section \ref{sec:walker_simulation_experiment}).
 Each realization of the topography yielded a slightly different Faraday threshold, which on average was \reply{$\gamma_\mathrm{F}=3.880g \pm 0.005g$}, where the interval represents the standard deviation. Repeating the simulations at six different memories \reply{($\gamma/\gamma_\mathrm{F}=77\%$, 86\%, 90\%, 94\%,} and 97\%) required readjusting the impact phase \reply{with fine tuned changes specific to each topography} \reply{($\theta_I/\pi=1.340 \pm0.001$, $1.392 \pm0.001$, $1.408 \pm0.001$, $1.421 \pm0.002$, $1.429 \pm0.002$\,rad, respectively)} to yield the same average speed, \reply{$v_0=5.13\pm0.15$\,mm\,s${}^{-1}$.} The results of these simulations are presented in Fig.\,\ref{fig:main_Fig_2} in the main text. The trajectories were unwrapped due to  crossings of the periodic domain. At high memory, the crossings are very infrequent due to the walker localization. At low memory, the walkers cross the domain frequently, which materializes in the linear dependence on time of the MSD. This behavior is rationalized through the diminished influence of the submerged topography on the wave field, which is less extensive and decays quickly in time, making the walker behavior approach that of a waveless particle  
 (see section \ref{sec:wave_field}, Fig.\,\ref{fig:sup_fig_LMwave}). 

 % REMAINDER
 %The lack of localization at low memory may thus not be attributed to the size of the domain, which is large relative to the extension of the wave field. Increasing the domain won't change anything, only probide more chances to generate a local arrangement that can trap the droplet. This can be checked through more simulations with the same domain size as now.

\subsubsection{Ergodicity \label{sec:ergodicity_check}}

\begin{figure}[]
\includegraphics[scale=1]{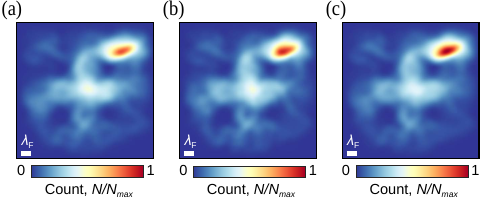}
\caption{\textbf{Ergodicity of the walker localization.} Position histogram arising from simulations of the walker model with
(a) 36 walker trajectories, each simulated up to 21,000 Faraday periods,  (b) 1 walker trajectory simulated up to 189,000 Faraday periods, and (c) 4 walker trajectories, each simulated up to 189,000 Faraday periods. In all cases, the emergent position histogram is virtually the same. 
\label{fig:sup_fig_ergodicity}}
\end{figure}

In order to establish the ergodicity of walker localization, we conducted a comparison between statistical results derived from a substantial number of relatively short simulations and those obtained from a few extended simulations. We recall that the histogram presented in Fig.\,\ref{fig:sup_fig_waveless_experiment_simulations}\fp{c} is the result of simulating 36 randomly initialized walkers, each integrated over 21,000 Faraday periods (see section \ref{sec:walker_simulation_experiment}). We compared the results to those obtained from 1 walker, simulated up to 189,000 Faraday periods, and 4 walkers, each simulated up to 189,000 Faraday periods.  The emergent position histograms are virtually the same (Fig.\,\ref{fig:sup_fig_ergodicity}).

\subsubsection{Base depth, droplet speed and tile width}

We performed tests to understand the role of the base depth, $H_0$,  walker average speed, $v_0$, and tile width, $W$. Reducing the base depth increases the influence of the submerged topography on the walker dynamics. Specifically, as the liquid layer becomes shallower, the random topography dominates more strongly the walker dynamics. The droplet thus explores a smaller fraction of the domain, thereby giving rise to a more pronounced peak the in the position histogram. In the opposite limit, as the layer depth increases, the heterogeneous topography distorts the droplet motion less, and eventually becomes imperceptible in the deep-fluid limit, $H_0\gtrsim \lambda_\mathrm{F}$, when the walker dynamics become ballistic. The walker characteristic speed, which may be tuned through the impact phase $\theta_I$, plays an equivalent role: the trajectory of faster droplets is less influenced by the submerged topography, while slower droplets are more susceptible to localization. Finally, observation of walker localization also requires the tile width to be comparable to the Faraday wavelength, $W\sim\lambda_\mathrm{F}$. When the tile width is small relative to the Faraday wavelength, $W\ll\lambda_\mathrm{F}$, the small-scale features of the topography become imperceptible to the walker, and so its dynamics becomes ballistic. When the tile width is large relative to the Faraday wavelength, $W\gg\lambda_\mathrm{F}$, the walker exhibits ballistic motion above the tiles. However, the droplet experiences deflections between the tiles, leading to diffusive motion in the long-time limit. The walker behavior thus become similar that expected for a particle devoid of wave-like characteristics.

%Another crucial parameter that affects the localization of the walker is the average depth, $h_{\text{mean}},$ which corresponds to the initial energy, $E_0$, in the Schr\"odinger simulations. Anderson localization as a quantum phenomenon present in the weak disorder limit, so one expects weaker localization as the average depth, $h_{\text{mean}}$, is increased. The original walker histogram is computed in the regime $h_{\text{mean}} = 1.85\, \text{mm}$, and by collecting walker statistics for $1.55\,\text{mm} \leq h_{\text{mean}} \leq 2.1\,\text{mm}$. The walker histograms (shown in Figure 3) show a negative correlation between average depth and localization strength, with deeper topographies exhibiting more diffusive behavior.[[Figure 3: Depth Variation]]

\subsubsection{Numerical convergence \label{sec:simulaitons_numerical_convergence}}

\begin{figure}[]
\includegraphics[scale=1]{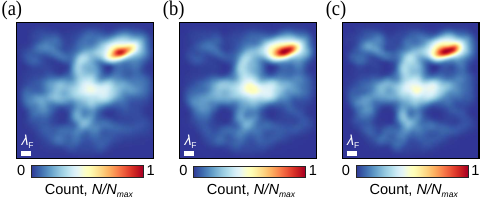}
\caption{\textbf{Numerical discretization of the walker simulations.} Position histograms  obtained with different spatio-temporal resolutions, specifically (a) $N\times N= 148\times 148$ ($\sim$8 points per tile), (b) $204\times 204$ ($\sim$11 points per tile), and  (c) $408\times 408$ ($\sim$21 points per tile). The temporal resolution was $\Delta t=T_\mathrm{F} L/(10 N\lambda_\mathrm{F})$ in (a,b), and $\Delta t=T_\mathrm{F} L/(20 N\lambda_\mathrm{F})$ in (c). 
\label{fig:sup_fig_mesh}}
\end{figure}

We performed numerical tests  to ensure that the spatio-temporal resolution was adequate to render discretization-independent results. Specifically, we repeated the simulations described in section \ref{sec:walker_simulation_experiment} with a coarser ($N\times N= 148\times 148$, $\Delta t=T_\mathrm{F} L/(10 N\lambda_\mathrm{F})$) and a finer ($N\times N= 408\times 408$, $\Delta t=T_\mathrm{F} L/(20 N\lambda_\mathrm{F})$) spatio-temporal discretization to compare the results with the reference case ($N\times N= 204\times 204$, $\Delta t=T_\mathrm{F} L/(10 N\lambda_\mathrm{F})$). The three simulations revealed similar localized statistics (Fig.\,\ref{fig:sup_fig_mesh}). We thus conclude that the intermediate resolution chosen for the reference case is adequate.

% ----------------------------------------------
%              WAVELESS PARTICLE
% ----------------------------------------------
\subsection{Waveless particles\label{sec:SI_waveless_simulations}}

To highlight the distinct behavior of the walkers, we compared in Fig.\,\ref{fig:main_Fig_1} their statistics to those arising from a `waveless' particle, with mass $m$ and average speed $v_0$, that evolves classically in a random background with the same form as the potential induced by the submerged bottom topography on the walkers. The problem may be non-dimensionalized using $x_c=W$, $t_c=W/v_0$ and $V_c=mW^2/v_0^2$ as the characteristic scales for length, time, and energy, respectively, where $W$ is the tile width of the random potential. The dimensionless governing equation for the particle position, $ {\bm{x}}_\mathrm{p}(t)$, then becomes
\begin{equation}    
\ddot{\bm{x}}_\mathrm{p}(t)=-\nabla V(\bm{x}_\mathrm{p}),
\label{eq:waveless_motion}
\end{equation}
which is initialized at the average speed, i.e. $ |\dot{\bm{x}}_\mathrm{p}(0)|=1$, in a random direction. The initial position is unimportant as long as its associated potential energy is the same as the mean value, thus ensuring that the particle's average speed remains close to unity. Alternatively, we may select the initial position randomly and readjust the initial speed to maintain the same total energy, and so obtain the same average speed. 
We select a potential with the same form as that induced by the walker's bottom topography in the experiments. We thus proceed as in section \ref{sec:localized_eigenfunctions}, and define 
\begin{equation}
V(\bm{x})=-K  H(\bm{x}),
\label{eq:dimless_potential_waveless}
\end{equation}
where, once again, $K=\kappa/V_c$ and $H(\bm{x})$ have dimensions of 1/length and length, respectively. Here, $H(\bm{x})$ is the random liquid depth distribution used in the experiments, and $\kappa$ the dimensional scaling constant presented in the main text, which has dimensions of energy/length. We note that subtracting the maximum depth as in \eqref{eq:dimless_potential} is not important in this case due to the gradient in \eqref{eq:waveless_motion}. The re-scaling constant $K$ was chosen to ensure the particle energy was greater than the maximum of the potential.

The random region was surrounded by a confining potential (corresponding to the wave damper with double the height about the base depth as in the experiment) to prevent the particle from escaping the domain of interest. The equation of motion was integrated using a fourth-order Runge-Kutta method with adaptive time stepping (Matlab ode45). \replyfinal{We ensured that the numerical tolerances were sufficiently stringent  so that the energy of the particles did not vary significantly throughout the simulations, typically remaining within $<1\%$ of the initial energy.}
A low-pass filter \reply{of radius (standard deviation)
$\sigma = 111 \%$ of the lattice frequency} was applied to the potential spectrum to prevent diverging gradients between tiles. \reply{Tests showed that varying the filter radius does not significantly influence the position histogram, where the essential feature is the absence of localization.} 
We matched the number of trajectories and their duration to those from experiments, \reply{ and conducted a series of simulations to verify that the particle dynamics converged to the same histogram, regardless of the initial conditions.}
For the simulations presented in Fig.\,\ref{fig:main_Fig_1}, the potential's re-scaling constant was $K=1.2\,\mathrm{mm}^{-1}$, and the ratio of total energy to maximum potential energy  $(\frac{1}{2}|\dot{\bm{x}}_\mathrm{p}|^2+V)/\max V = 1.32$.

%\begin{figure*}
%\includegraphics[scale=1]{SIFigs/SFigure2_lite_PRX.pdf}
%\caption{\textbf{Emergent localized statistics in the walker system.} Comparison of the speed-colored trajectory (first row), position histogram (second and third rows), average speed map (fourth row) for (a) a waveless particle (see section \ref{sec:SI_waveless_simulations}), and a walker according to (b) experiments (see section \ref{sec:SI_experiment}), and (c) simulations (see section \ref{sec:SI_walker_simulation}). The position histograms, $N(\bm{x})$ (computed using 3 bins per wavelength in each direction), were first normalized such that $\iint_\Omega N(\bm{x}) \,\mathrm{d}S=1$, and later re-scaled by the maximum of the three histograms, $N_\mathrm{max}$. The average speed maps were computed using a $78\times78$ regular grid and a Gaussian-weighted average, with standard deviation $\sigma=\lambda_\mathrm{F}/4$, applied to data points within $\lambda_\mathrm{F}/4$ distance away from each grid point.
%\label{fig:sup_fig_waveless_experiment_simulations}}
%\end{figure*}

%\clearpage

%\onecolumngrid

\def\bibsection{\section*{REFERENCES}} 

% Biliography
%\bibliography{refs_SI}
% Biliography
%\bibliography{refs_HSLs,refs_Anderson,BeamsBib}
\bibliography{refs_Anderson}

\end{document}